\newif\iflong
\newif\ifshort

\longtrue

\iflong 
\else
\shorttrue
\fi

\documentclass{article}
\usepackage{ijcai21}

\usepackage{times}

\usepackage{soul}
\usepackage{url}

\usepackage[pagebackref]{hyperref}
\hypersetup{%
  backref=true, 
  pagebackref=true, 
  hypertexnames=true,
  colorlinks=true,citecolor=green!60!black,linkcolor=red!60!black%
}
\usepackage[textsize=tiny,textwidth=1.5cm,linecolor=green!70!black, backgroundcolor=green!10, bordercolor=black,disable]{todonotes}
\usepackage[utf8]{inputenc}
\usepackage[small]{caption}
\usepackage{graphicx}
\usepackage{amsmath}
\usepackage{booktabs}
\usepackage{paralist}
\usepackage{amssymb,amsthm,dsfont}
\usepackage{color}
\usepackage{xspace}
\usepackage{enumerate}
\usepackage{mathtools}
\usepackage{bbm}
\usepackage{comment}
\usepackage{paralist}
\usepackage{nicefrac}
\usepackage{multirow}
\usepackage{wrapfig}

\usepackage{subcaption}
\usepackage[noend,ruled,linesnumbered]{algorithm2e}

\SetCommentSty{mycommfont}
\usepackage{xifthen}

\usepackage{scrextend}

\usepackage{needspace}

\usepackage{nicefrac}

\newcommand{\newH}[1]{#1}

\newcommand{\old}[1]{}

\usepackage[absolute]{textpos}  %

\newcommand{\myparagraph}[1]{\smallskip
  
\noindent  \textbf{#1}}

\newcommand{\myemph}[1]{{\color{darkgreen!70!black}\emph{#1}}}
\newtheorem{theorem}{Theorem}[section]
\newtheorem{corollary}[theorem]{Corollary}
\newtheorem{example}{Example}
\newtheorem{lemma}[theorem]{Lemma}

\newtheorem{proposition}[theorem]{Proposition}
\newtheorem{claim}{Claim}

\theoremstyle{remark}

\theoremstyle{definition}
\newtheorem{definition}{Definition}

\newcommand{\probname}[1]{\textsc{#1}}

\usepackage{xcolor}

\newcommand{\decprob}[3]{
   \begin{center}%
    \begin{minipage}{0.92\linewidth}%
      \textsc{#1}\\[0.2ex]
      \textbf{Input:} #2\\[0.2ex]
      \textbf{Question:} #3
    \end{minipage}%
  \end{center}
}

\definecolor{winered}{rgb}{0.6,0.1,0.1}
\definecolor{darkblue}{rgb}{0,0,0.4}
\definecolor{darkgreen}{rgb}{0.01,0.6,0.1}

\usepackage{tikz}
\usetikzlibrary{calc,arrows,shapes,shapes.geometric,matrix,decorations.pathreplacing,positioning,patterns,backgrounds,fit,topaths,mindmap}

\tikzset{matrixsc/.style={matrix of math nodes, ampersand replacement=\&, row sep=-7pt, column sep=-4pt}}

\makeatletter
\newcommand{\gettikzxy}[3]{%
  \tikz@scan@one@point\pgfutil@firstofone#1\relax
  \edef#2{\the\pgf@x}%
  \edef#3{\the\pgf@y}%
}
\makeatother

\newcommand{\twoqsat}{\textsc{$\exists\forall$SAT}\xspace}
\newcommand{\notoneinthree}{\textsc{Not-1-in-3-$\exists\forall$3SAT}\xspace}
\newcommand{\maxtruesat}{\textsc{Max-True-3SAT-Compare}\xspace}
\newcommand{\knapsack}{\textsc{Knapsack}\xspace}

\newcommand{\sigmatwop}{$\Sigma_2^{\text{P}}$}
\newcommand{\parallelnp}{$\text{P}_{||}^{\text{NP}}$}

\newcommand{\maxone}{\ensuremath{\mathsf{max}\text{-}\mathds{1}}}

\newcommand{\true}{\mathsf{true}}
\newcommand{\false}{\mathsf{false}}

\newcommand{\zero}{\boldsymbol{0}}

\newcommand{\lit}{\mathsf{lit}}

\newcommand{\cost}{\ensuremath{\mathsf{c}}}
\newcommand{\lowervec}{\ensuremath{\boldsymbol{\ell}}}
\newcommand{\uppervec}{\ensuremath{\boldsymbol{u}}}

\newcommand{\contvec}{\ensuremath{\boldsymbol{b}}}
\newcommand{\typevec}{\ensuremath{{\boldsymbol{\tau}}}}
\newcommand{\sat}{\ensuremath{\mathsf{sat}}}
\newcommand{\util}{\ensuremath{{\mu}}}
\newcommand{\utilR}{\ensuremath{\util^{\utilmode}}}

\newcommand{\probWinner}{\probname{$\PBrule$-Winner}\xspace}
\newcommand{\probNotWinner}{\probname{co-$\PBrule$-Winner}\xspace}
\newcommand{\probNotWinnerR}[1]{\probname{co-#1-Winner}\xspace}
\newcommand{\probWinnerR}[1]{\probname{#1-Winner}\xspace}
\newcommand{\probSpend}{\probname{$\PBrule$-Donation}\xspace}
\newcommand{\probSpendR}[1]{\probname{#1-Donation}\xspace}

\newcommand{\PBrule}{R}
\newcommand{\agggreedy}{\ensuremath{\mathcal{G}}}
\newcommand{\greedset}{\ensuremath{\mathbb{G}}}
\newcommand{\Sc}{\agg}
\newcommand{\ScR}{\agg^{\utilmode}_{\aggmode}}
\newcommand{\RR}{\PBrule^{\utilmode}_{\aggmode}}

\newcommand{\agg}{\mathsf{score}}
\newcommand{\FB}{\ensuremath\mathbb{C}(I)}
\newcommand{\pb}{\textsf{PB}\xspace}

\newcommand{\pbinstafter}{\ensuremath{B,\lowervec,\uppervec}}
\newcommand{\score}{\ensuremath{s}}

\newcommand{\don}{\ensuremath{\delta}}
\newcommand{\Pot}{\ensuremath{\mathcal{P}}}
\newcommand{\profileinstance}{\ensuremath{(t,(\cost_j)_{j\in[m]},(\typevec_j)_{j\in [m]}, (\sat_i)_{i\in[n]},(\contvec_i)_{i\in[n]})}}
\newcommand{\aggsum}{\ensuremath{\agg_{\Sigma}}}
\newcommand{\aggmin}{\ensuremath{\agg_{\min}}}

\newcommand{\aggSet}{\ensuremath{\mathcal{R}}}

\newcommand{\zeros}{\ensuremath{0^{\enn+1}}}

\newcommand{\enn}{\ensuremath{{\hat{n}}}}
\newcommand{\emm}{\ensuremath{{\hat{m}}}}

\newcommand{\claimqed}{$\blacklozenge$}

\newcommand{\utilmode}{\ensuremath{\star}}
\newcommand{\aggmode}{\ensuremath{\diamond}}

\newcommand{\mypa}[1]{
  \smallskip
  \noindent \underline{\emph{#1}}
  \smallskip}

\DeclareMathOperator*{\vx}{vx}
\DeclareMathOperator*{\vy}{vy}

\newcommand{\dptime}{\ensuremath{n\cdot m +  (B+1)\cdot (m+1)^{t+1}\cdot t}}

\newcommand{\mythm}[1]{[T~\ref{#1}]}

\DeclareMathOperator*{\argmax}{arg\,max}

\usepackage[noend,ruled,linesnumbered]{algorithm2e}

\usepackage[sort&compress,nameinlink,capitalize]{cleveref}
\crefname{figure}{Figure}{Figures}
\crefname{algorithm}{Algorithm}{Algorithms}
\crefname{proposition}{Proposition}{Propositions}
\crefname{theorem}{Theorem}{Theorems}
\crefname{claim}{Claim}{Claims}
\crefname{lemma}{Lemma}{Lemmas}
\crefalias{AlgoLine}{line}

\definecolor{OKgreen}{HTML}{6B9B00}
\definecolor{NOred}{HTML}{83004F}
\usepackage{pifont}%
\newcommand*\OK{\textcolor{OKgreen}{\ding{51}}}
\newcommand*\NO{\textcolor{NOred}{$\times$}}

\usepackage{etoolbox} %

\newcommand{\appsymb}{$\star$}
\newcommand{\appref}[1]{{\hyperref[#1]{\appsymb}}}

\newcommand{\appendixsection}[1]{%
}

\ifshort 
\title{Participatory Budgeting with Donations and Diversity Constraints}
\else
\title{%
  Participatory Budgeting with Donations and Diversity Constraints}
\fi
\author{
  Jiehua Chen$^1$ \and Martin Lackner$^1$ \And Jan Maly$^1$
\affiliations
$^1$TU Wien, Vienna, Austria\\
\emails
jiehua.chen@tuwien.ac.at, \{lackner,jmaly\}@dbai.tuwien.ac.at
}

\begin{document}

\maketitle

\begin{abstract}
Participatory budgeting (PB) is a democratic process where citizens jointly decide on how to allocate public funds to indivisible projects. %
 This paper focuses on PB processes
where citizens may give additional money to projects they want to see funded.
We introduce a formal framework for this kind of PB with donations.
 Our framework also allows for diversity constraints, meaning that each project belongs to one or more types, and there are lower and upper bounds on the number of projects of the same type that can be funded.
  We propose three general classes of methods for aggregating the citizens' preferences in the presence of donations and analyze their axiomatic properties.
  Furthermore, we investigate the computational complexity of determining the outcome of a PB process with donations and of finding a citizen's optimal donation strategy.
\end{abstract}

\section{Introduction}

Participatory budgeting (PB) \cite{Caba04,Shah07} is a democratic tool 
that enables voters to directly decide about budget spending.
The general procedure of PB is that voters are presented a number of %
projects (e.g., building a library or a park) and are asked to vote on these projects.
Then, a PB~aggregation rule is used to select a subset of projects---a so-called bundle---to be funded.
This bundle has to be feasible, which typically means that the total cost must not exceed the available budget, and sometimes further adhere to fairness constraints.

In this paper, we study participatory budgeting with donations. In our model, voters can pledge donations for projects they support. If such a project is funded, the donations are levied and only the remaining cost is paid by the public budget. Consequently, projects with donations can be funded with a reduced impact on the public budget.
At first glance, allowing donations in PB referenda brings a major advantage:
as the total available budget increases, a larger overall satisfaction is achievable.
In addition, voters with an intense preference for a project can support this project financially and thus increase the chance of it being funded.

Our chosen model is based on PB with cardinal preferences, i.e., voters have numbers associated with projects that reflect their preferences. Cardinal preferences capture, e.g., settings with approval ballots (only 0 and 1 are used), settings where voters can distribute points to projects (where usually the sum of points is bounded), and settings where these numbers accurately correspond to the utility of voters.
Further, we allow for diversity constraints \cite{BFILS-diversity,BenabbouCZ19,YangGS19,ChenGanianHamm2020ijcai-diversestable}:
Each project belongs to one or more types (based on classifications such as ``youth and education'' or ``transport and mobility'')
and for each type there is a minimum and maximum number of projects to be funded. %
This can also model city-wide referenda where districts have their own ``project quota''.
(It is straightforward to also include constraints with a minimum/maximum amount of budget spent, cf. the work of \citeauthor{HKPP21aaai}~\citeyear{HKPP21aaai}.)
The goal of this paper is to mathematically analyze PB with donations and to investigate the computational complexity of problems arising therein.
In general, it is far from obvious that donations in PB are desirable at all.
Thus, we define a number of desiderata for PB~aggregation %
rules that take donations into account:
\begin{compactenum}[D1]
  \item\label{ax:no-harm} \myemph{Donation-no-harm}. {Allowing donations \emph{should not} make any voter less satisfied (independent of whether the voter donated herself).}
  \item\label{ax:proj-mono} \myemph{Donation-project-monotonicity}. {Increasing the donation from any voter to a \emph{winning} project \emph{should not} lead to this project not winning anymore.}

  \item\label{ax:welfare-mono} \myemph{Donation-welfare-monotonicity}. {Increasing the donation from any voter to a project \emph{should not} lead to a decrease of the %
    social welfare~(for a given welfare definition).}
  \item\label{ax:voter-mono} \myemph{Donation-voter-monotonicity}. {Donating to a project \emph{should not} make a voter less satisfied than not donating to this project (keeping her donations to other projects unchanged).}
\end{compactenum} 

In this paper, we exemplarily consider four standard PB aggregation rules, all of which are optimization methods that select an ``optimal'', feasible bundle of projects (we write $\PBrule\in\aggSet$\todo{One reviewer mentions that this appears a bit early.} to denote these four rules, see Section~\ref{sec:aggregation-methods}).
A natural approach for handling donations is to directly apply $\PBrule$; this approach takes donations into account since reduced project costs due to donations lead to more (and larger) feasible bundles.
However, as we show in this paper, this approach violates D\ref{ax:no-harm},
i.e., introducing donations into a PB process may disadvantage some voters;
we see this as a major downside of this approach.

Consequently, we propose two further approaches how a PB aggregation rule~$\PBrule$ can be modified to take donations into account.
Sequential-$\PBrule$ first applies $\PBrule$ ignoring all donations. If after the application of $\PBrule$ some budget is left (due to donations for the funded projects), $\PBrule$ is applied (again ignoring all donations) with the remaining budget; this step is repeated if necessary.
Only in a last step, $\PBrule$ is applied with project costs reduced by donations; this step guarantees that an exhaustive (maximal) set of projects is selected.
We can show that Sequential-$\PBrule$ satisfies desiderata D\ref{ax:no-harm} and D\ref{ax:proj-mono} for all considered PB~aggregation rules $\PBrule\in\aggSet$, but it fails D\ref{ax:welfare-mono} and D\ref{ax:voter-mono}.

The third approach, Pareto-$\PBrule$, first applies $\PBrule$ ignoring all donations.
Based on the winning bundle, it selects a Pareto-optimal improvement of this bundle taking donations into account.
When choosing a notion of Pareto improvement that is compatible with the notion of welfare used in D\ref{ax:welfare-mono},
we can prove that Pareto-$\PBrule$ satisfies D\ref{ax:welfare-mono}. %
It also satisfies  D\ref{ax:no-harm} and D\ref{ax:proj-mono}, but fails D\ref{ax:voter-mono}.

As a final result in our axiomatic analysis, we show that it is no coincidence that all three proposed principles fail D\ref{ax:voter-mono}: we prove that---under reasonable assumptions---D\ref{ax:voter-mono} is impossible to satisfy.
All axiomatic results hold independent of diversity constraints and are succinctly summarized in Table~\ref{tab:desiderata}.

\begin{table}
  \caption{Desiderata for the PB rules and complexity results.}\label{tab:desiderata}
  \centering
  \begin{tabular}{@{}l@{\,\;}c@{\,\;}c@{\,\;}c@{\,\;}cc@{\;}c@{}}
    \toprule
    Rule~$\PBrule$        & D1    & D2    & D3   &  D4   & \probWinner &  \probSpend\\
    \midrule
    Apply~$\PBrule$
                                       &  \NO  & \OK   & \OK  &  \NO  & coNP-c & \sigmatwop-c \\
    Sequential-$\PBrule$               &  \OK  & \OK   & \NO  &  \NO   & coNP-h & \sigmatwop-c \\
    Pareto-$\PBrule$                   &  \OK  & \OK   & \OK  &  \NO   & coNP-h & \sigmatwop-c \\
    \bottomrule
   \end{tabular}
 \end{table}

In addition to the axiomatic analysis, we study the computational complexity
questions that arise in our framework.
We focus on two computational problems, one on winner determination and the other on effective donation.

The first problem, called \probWinner, is to decide whether a given bundle is a winner under~$\PBrule$.
We show that for $\PBrule^{+}_\Sigma$ and for a constant number of types, the problem can be solved in pseudo-polynomial time and is weakly coNP-hard.
For the three other rules, we prove coNP-completeness.
The hardness results also hold for the Sequential-$\PBrule$ and Pareto-$\PBrule$ variants.

The second problem, called \probSpend, is to decide whether a given voter can effectively spend a given amount of money so as to achieve a higher utility (than with the initial donation).
While it is straight-forward to see that the problem is naturally contained in \sigmatwop, a complexity class from the second level of the polynomial hierarchy~\cite{Pap94},
the power of the diversity constraints enable us to show that it is indeed \sigmatwop-hard, even under severe restrictions to the input instances. 
In particular, this implies that the problem is not easily amenable to SAT or ILP solvers.
We also show a somewhat unexpected result that even if no diversity constraints are imposed, \probSpend remains hard for the complexity class~\parallelnp~\cite{Pap94} (except for $\PBrule^{+}_\Sigma$, which aims to find a feasible bundle with maximum sum of overall satisfaction).
For $\PBrule^{+}_\Sigma$, finding an effective donation is at least beyond NP.

To sum up, our work provides a first analysis of PB with donations. We discuss features and pitfalls of this idea, propose methods to handle donations, and analyze their computational demands.

Due to space limits, most proofs had to be omitted; full proofs can be found in the appendix. 

\paragraph{Related work.}
Participatory budgeting has received substantial attention
through the lens of (computational) social choice in recent years, see e.g.\ \cite{FGM16,ALT18,FPPV19,GKSA19,Laru21}; we refer to the survey by \citeauthor{AzSh20}~\shortcite{AzSh20} for a detailed overview of this line of research.
However, donations have not been considered in the
indivisible PB model that we are concerned with in this work.
The allocation of donations has been studied in a model related to 
divisible participatory budgeting
albeit without external budget \cite{BraBraPetStrSuk2020}.

In contrast, diversity constrains have been studied 
in PB in the form of an upper bound on the
amount of money spent on each type \cite{JSTZ20arxiv} . However, to the best of our knowledge,
our work is the first to consider diversity constrains with both upper and lower bounds.
Additionally, PB with project interactions \cite{JST20ijcai} is another approach 
using project types to guarantee diverse outcomes, albeit by changing the
utility functions of the voters instead of the set of feasible outcomes.
Finally, diversity constraints have been studied in multi-winner voting 
\cite{BFILS-diversity,celis2018multiwinner,YangW18,BeiLPW20},
which can be considered a special case of PB 
where projects have unit costs.

\section{Preliminaries}
Given a non-negative integer~$z$, we use \myemph{$[z]$} to denote the set~$\{1,2,\ldots,z\}$.
Given a vector~$\boldsymbol{x}\in \mathds{Z}^k$ of dimension~$k$, we use \myemph{$\sum{\boldsymbol{x}}$} to denote the sum of the values in~$\boldsymbol{x}$\iflong, i.e., $\sum{\boldsymbol{x}} = \sum_{j\in [k]}\boldsymbol{x}[j]$\fi.

The input of our participatory budgeting problem consists of a set of $m$~projects $C=[m]$, a set of $n$ voters~$V=[n]$ and a set of types~$T=[t]$ along with the following extra information:

\noindent Each project~$j\in C$ has 
\iflong
\begin{enumerate}[(i)]
        \item \fi a \myemph{cost~$\cost_j\in \mathds{N}$}, and
\iflong \item \fi a \myemph{type vector~$\typevec_{j}\in \{0,1\}^t$}, where $\typevec_j[z]=1$ means that project~$j$ has type~$z$.
\iflong
\end{enumerate}
\fi
\iflong \noindent \fi Each voter~$i\in V$ has
\iflong \begin{enumerate}[(i)]
  \item \fi \ifshort (i) \fi a \myemph{satisfaction function~$\sat_i\colon C \to \mathds{N}_0$}, which models how much she would like a project to be funded%
, \iflong where %
  for each object~$j\in C$ it holds that $\sat_i(j) > 0$ if and only if voter~$i$ finds $j$ \myemph{acceptable} and should be funded, \fi and
\iflong  \item \fi \ifshort (ii) \fi a \myemph{contribution vector~$\contvec_i\in \mathds{N}^m$} such that for each project~$j\in C$ the value~$\contvec_i[j]$ indicates how much money she is willing to donate if project~$j$ should be selected.
\iflong
\end{enumerate}
\fi

We call $\Pot=\profileinstance$ a \myemph{\pb{} profile} and each possible subset of projects a \myemph{bundle}.
A \pb{} instance $I=(\Pot, \pbinstafter)$ contains, in addition to the \pb{}
profile $\Pot$, a set of constraints that a winning bundle has to satisfy.
These are determined by the budget~$B \in \mathbb{N}$ and the diversity constraints
specified by two vectors $\lowervec \in \mathbb{N}^t$ and $\uppervec \in \mathbb{N}^t$
representing the lower and upper bound on the number of projects funded per type. 
Throughout the paper, we assume that~$\Pot$ denotes a \pb{} profile of the form~$\profileinstance$ 
and $I$ denotes a \pb{} instance of the form $(\Pot, \pbinstafter)$.

We say that a bundle~$A\subseteq C$ is \myemph{feasible} for~$I$ %
if both the budget and diversity constraints are fulfilled, i.e., if:
\begin{description}%
  \item[Budget constraint:]
  {$\sum_{j\in A} \max(0,\cost_j-\sum_{i\in V}\contvec_i[j]) \le B$.}
  \item[Diversity constr.:] \hfill $\lowervec[z] \le \sum_{j\in A}\typevec_j[z] \le \uppervec[z]$,\quad $\forall z\in T$. %
\end{description}
We write $\mathbb{C}(I)$ to denote the set of all feasible bundles for~$I$.
We say that $A$ is \myemph{exhaustive} if %
adding any additional project to~$A$ will violate the budget or diversity constraints\iflong, i.e.,
for each $j\in C\setminus A$ it holds that $A\cup\{j\}$ is not feasible\fi.

Finally, we introduce some additional notions and notations.
We say that voter~$i$'s contribution vector $\contvec_i$ is \myemph{satisfaction consistent} if
\iflong \begin{compactenum}[(i)]
  \else \begin{inparaenum}[(i)]
    \fi
    \item %
    voter~$i$ only donates to projects for which she has positive satisfaction value, i.e., for all~$j\in C$ with $\sat_i(j)=0$ it holds that $\contvec_i[j]=0$, and
    \item\label{donation-satisfaction-consistent} for each two projects~$j$ and $j'$ with $\sat_i(j) > _i \sat_i(j')$ it holds that $\contvec_i[j]> \contvec_i[j']$.
\iflong
\end{compactenum}
\else
\end{inparaenum}
\fi
Further, we say that a contribution vector~$\contvec'$ is a \myemph{$j$-variant of a contribution vector~$\contvec$} if
for each $j'\in C$ with $j'\neq j$ it holds that $\contvec'[j']=\contvec[j']$ (they only differ for project~$j$).
Given a contribution vector~$\contvec'_v$ for a voter~$v$ we use \myemph{$I-\contvec_v+\contvec'_v$}
to denote the \pb{} instance where the contribution vector of $v$ is replaced with $\contvec'_v$.
\iflong
Formally, $I-\contvec_v+\contvec'_v \coloneqq (\Pot', \pbinstafter)$, where $P'\coloneqq ((\cost_j)_{j\in[m]},(\typevec)_{j\in [m]}, (\sat_i)_{i\in[n]},(\contvec_1,\contvec_2,\ldots, \contvec_{v-1}$, $\contvec'_v,\contvec_{v+1},\ldots, \contvec_{n})_)$.
\fi
For a \pb{} instance~$I$,
let \myemph{$I^0$} denote the \pb{} instance which differs from~$I$
only in that all donations are zero.

\section{Aggregation Methods}\label{sec:aggregation-methods}

In order to select the feasible bundle that offers the highest satisfaction 
to the voters, we aggregate the satisfactions in two steps.
First, we define the voters' utility towards a bundle of projects, and then we aggregate the utilities of all voters. We consider two options for each step, respectively.

\iflong
\subsection{Lifting satisfaction functions to bundles}\label{subsec:utility-functions}
\fi

We call a function which lifts satisfaction functions for single projects to bundles 
\myemph{utility functions} and write $\util$ for a utility function.
\iflong Among the many possible utility functions, 
we
\fi
consider two standard functions which are suited for cardinal preferences: summing the satisfaction of each project in the bundle (additive) or choosing the highest satisfaction of all projects bundle (maximum).
\iflong
Let $I$ be a \pb{} instance and~$A\subseteq C$ a bundle.
\fi
\begin{align*}
\util_i^{+}(I,A) &\coloneqq \sum_{c\in A}\sat_{i}(c) \tag{additive}\\
\util_i^{\max}(I, A) &\coloneqq \max\limits_{c\in A}\sat_i(c) \tag{maximum}
\end{align*}

We will omit the first input parameter~$I$ from the function and write~$\util_i(A)$ instead if it is clear from the context which profile we are referring to.

\iflong
\subsection{PB aggregation rules}
\fi

Next, given a \pb{} instance~$I$ and a bundle~$A\in \mathbb{C}(I)$, a \myemph{scoring function}~$\agg$ computes a number indicating the overall utilities of the voters towards~$A$.
An aggregation rule based on~$\agg$ will return a feasible bundle~$A$ among all feasible ones with maximum~$\agg(A)$. As convention, if there are multiple feasible bundles with maximum score, then we select one according to an arbitrary but fixed tie-breaking rule.
We consider two types of scoring functions:
\iflong
\paragraph{Sum.}
This type of scoring functions returns the sum of utilities of the voters towards a given bundle.
Since we are interested in two utility functions~$\util^{\utilmode}\in \{\util^{\max}, \util^{+}\}$, there are two scoring functions of this type, defined as follows:
For each $\star\in \{\max, +\}$, we define
\begin{align*}
 \agg_\Sigma^{\utilmode}(I,A) \coloneqq \sum_{i\in V}\util_i^{\star}(P,A).
\end{align*}

\iflong\noindent The aggregation rule~{$\PBrule_\Sigma^{\utilmode}$} selects a feasible bundle~$A$ with maximum~$\agg_\Sigma^{\utilmode}(I,A)$, i.e., \myemph{$\PBrule_\Sigma^{\utilmode}(I) = \argmax_{A\in \mathbb{C}(I)}\agg_\Sigma^{\utilmode}(I,A)$}. %
\fi  
%
%

\paragraph{Minimum.}
This type of scoring functions returns the minimum utility among all voters towards a given bundle.
For each $\utilmode\in \{\max, +\}$, we define
 \begin{align*}
    {\agg_{\min}^{\utilmode}(I,A)} \coloneqq \min_{i\in V}\util_i^{\utilmode}(A).
 \end{align*}

\noindent The aggregation rule~$\PBrule_{\min}^{\utilmode}$ selects a feasible bundle~$A$ with maximum~$\agg_{\min}^{\utilmode}(I,A)$, i.e., \myemph{$\PBrule_{\min}^{\utilmode}(I) = \argmax_{A\in \mathbb{C}(I)}\agg_{\min}^{\utilmode}(I,A)$}. 
\fi
\ifshort
the \myemph{sum} scoring functions return the sum of utilities of the voters towards a given bundle;
the \myemph{min} scoring functions return the minimum satisfaction.
For each $\utilmode\in \{\max, +\}$, we define
\begin{align*}
 \agg_\Sigma^{\utilmode}(I,A) &\coloneqq \sum_{i\in V}\util_i^{\utilmode}(P,A),\tag{sum}\\
 {\agg_{\min}^{\utilmode}(I,A)} &\coloneqq \min_{i\in V}\util_i^{\utilmode}(A).\tag{min}
\end{align*}
For~$\utilmode\in \{\max,+\}$ and $\aggmode\in \{\sum,\min\}$,
the aggregation rule~$\PBrule_{\aggmode}^{\utilmode}$ selects a feasible bundle~$A$ with maximum~$\agg_{\aggmode}^{\utilmode}(I,A)$.
\fi  
%
Thus, altogether, we look at four aggregation rules. We write $\aggSet\coloneqq \{\PBrule_{\min}^{+},\PBrule_{\min}^{\max},\PBrule_\Sigma^{+},\PBrule_\Sigma^{\max}\}$ for the set of these four.

\iflong
\subsection{Aggregation with donations}
\fi

As the four aggregation methods maximize a function over the set of feasible bundles,
they can simply handle donations via the definition of feasibility (cf.\ budget bound).
Note that, using this approach, the effect of a donation is equivalent to reducing the cost of the respective project.
However, as we will see, this simple way of handling donations has some
undesirable consequences. Therefore we also consider two other natural variants for handling donations.
Let $\PBrule\in\aggSet$.

\iflong
The first variant of~$\PBrule$, called \myemph{Sequential-$\PBrule$}, is to proceed~$\PBrule$ sequentially on the instance when no donations are considered:
In each turn a bundle according to $\PBrule$ is chosen (i.e., without consideration of donations).
If this bundle requires less budget than available (due to donations),
another round of~$\PBrule$ is run, again on the remaining projects without donations. This is repeated until no further projects can be afforded.
Finally, $\PBrule$ needs to be applied again, this time with consideration of the donations
to ensure that the final bundle is exhaustive.

The second variant, \myemph{Pareto-$\PBrule$} is to compute a bundle according to a \pb{} rule~$\PBrule$
(without considering donations) and then to select a Pareto-optimal improvement of this bundle
taking donations into account. Here, the Pareto-optimality is defined as follows:
A bundle~$A$ \myemph{Pareto-dominates} another bundle~$A'$ regarding a utility function~$\util$ if
\begin{compactitem}[$\bullet$]
  \item for each voter~$i\in V$ it holds that $\util_i(A)\ge \util_i(A')$
  and
  \item there exists a voter~$i\in V$ with $\util_i(A) > \util_i(A')$.
\end{compactitem}
In this case, we also say that $A$ \myemph{$\util$-dominates} $A'$.
A bundle~$A$ is \myemph{Pareto-optimal} with respect to~$\util$,
if no other \emph{feasible} bundle \myemph{$\util$-dominates}~$A$.

For a formal definition of the two methods, let $\util$ be a utility function,
let $\PBrule$ be an aggregation rule and $\agg$ the corresponding scoring function.
Then, we define for a \pb{} instance~$I=(\Pot, \pbinstafter)$ with
$\Pot=\profileinstance$ the following two variations of $\PBrule$:
\fi

%
\iflong \begin{compactitem}[--]
  \item \fi
  \myemph{Sequential-$\PBrule$} 
first applies $\PBrule$ for $\Pot^0$. If afterwards some budget is left (due to donations), $\PBrule$ is applied again with the remaining budget but still without donations; this step is repeated as long as new projects are added.
In a last step, $\PBrule$ is applied directly, thus guaranteeing an exhaustive bundle.
For details, see Algorithm~\ref{algo:seq-lambda}. Therein, we write
$\Pot(X)$ (for $X\subseteq C$) to denote the profile~$\Pot$ restricted to projects in~$X$.

\begin{algorithm}[t!] %
  $C \leftarrow [m]$; \tcp*{$[m]$ denotes the project set in the input}%
  
  \While{$C$ changes in the previous iteration}{
    $A_0 \leftarrow \PBrule(\Pot^0(C), \pbinstafter)$\;
    $C \leftarrow C \setminus A_0$\;
    $\lowervec \leftarrow \lowervec - \sum_{j\in A_0}\typevec_j $;\ \quad     
    $\uppervec \leftarrow \uppervec - \sum_{j\in A_0}\typevec_j $\;
    $B \leftarrow B - \sum\limits_{j \in A_0} \max(0,~~ \cost_j -\!\! \sum\limits_{j\in A_0,i\in V} \contvec_i[j])$\;
  }
  \Return $([m]\setminus C) \cup \PBrule( \Pot(C),\pbinstafter)$
\caption{Sequential-$\PBrule(I)$ with $I=(\Pot, \pbinstafter)$}\label{algo:seq-lambda}
\end{algorithm}

\iflong \item \fi The second method is \myemph{Pareto-$\PBrule$}.
\ifshort We say that $A\subseteq C$ \myemph{$\util$-dominates}~$A'\subseteq C$ if
for each voter~$i\in V$ it holds that $\util_i(A)\ge \util_i(A')$
and there exists a voter~$i\in V$ with $\util_i(A) > \util_i(A')$. \fi
Let $A_0 = \PBrule(I^0)$.
Now, consider the 
set of bundles~$\mathcal{X}$ containing \begin{inparaenum}[(i)] \item $A_0$ and \item all bundles $A^*\in\mathbb{C}(I)$ that $\util$-dominate $A_0$.   Pareto-$\PBrule$\ chooses a bundle~$A\in\mathcal{X}$ with maximum $\agg(A)$.
\end{inparaenum}
\iflong \end{compactitem} \fi

\begin{example}
Consider the following \pb instance~$I$ with $5$ projects~$p_1,\ldots,p_5$, two voters, budget~$B=5$, and without diversity constraints:
\begin{center}
\begin{tabular}{l*{8}{c}}
          & $c(\cdot)$ & $\sat_1$ & $\sat_2$  & $b_1$  & $b_2$  \\
\midrule                                                       
$p_1$     &   $3$      &      $5$ &    $5$    &  $1$   &   $0$  \\
$p_2$     &   $3$      &      $9$ &    $0$    &  $0$   &   $0$  \\
$p_3$     &   $2$      &      $1$ &    $2$    &  $0$   &   $0$  \\
$p_4$     &   $3$      &      $3$ &    $3$    &  $0$   &   $0$  \\
$p_5$     &   $1$      &      $1$ &    $1$    &  $0$   &   $0$  \\
\end{tabular}
\end{center}\todo{One reviewer: the instance is not satisfaction consistent.}
We consider rule~$\PBrule^{+}_\Sigma$ and its sequential- and Pareto-variants.
One can verify that the winner under $\PBrule^{+}_\Sigma$ is $A_1=\{p_1,p_2\}$ as it maximizes $\agg_\Sigma^+$ ($=19$).
Without donations, the winner is $A_0=\{p_1,p_3\}$.
Hence, Sequential-$\PBrule_\Sigma^+$ starts by selecting~$A_0$. 
Then it runs $\PBrule_\Sigma^+$ on the instance created by removing $p_1$ and $p_3$
from $\Pot^0$ with budget of $1$ (the cost for $p_1$ is reduced by $1$ due to voter~$1$'s donation).
Now, $\{p_5\}$ is the winner (for~$\PBrule^{+}_\Sigma$), leaving $0$ budget for the next round.
Hence the final winning bundle is $A_2=\{p_1,p_3,p_5\}$. 

Pareto-$\PBrule_\Sigma^+$ maximizes~$\agg_\Sigma^+$ among the projects which $\util^+$-dominate $A_0$. While $A_1$ has a higher score than $A_0$
it does not $\util^+$-dominate $A_0$ since voter~$2$ is worse off~($\util^{+}_{2}(A_1) = 5 < 7=\util^{+}_{2}(A_0)$). Indeed, $A_3=\{p_1,p_4\}$
and $A_2$ are the only feasible bundles which $\util^+$-dominate $A_0$.
Among those $A_3$ has the highest score and is hence the winner under~Pareto-$\PBrule_\Sigma^+$.
\end{example}

\section{Axioms Regarding Donations}

\looseness=-1

First, we note that applying any of our standard aggregation rules in the presence of donations
comes with the considerable disadvantage that some voters may end up less
 satisfied with the outcome than in a process without donations.

\begin{definition}[$\hat{\util}$-donation-no-harm]\label{def:no-harm}
  An aggregation rule $\PBrule$ satisfies \myemph{$\hat{\util}$-donation-no-harm} if for each \pb{} instance~$I$\todo{One reviewer is confused by the use of $\hat{\util}$ and just to just use $\util$. I think we should keep it.}\  
and each voter~$x$ it holds that
\[\hat{\util}_{x}(\PBrule(I)) \geq \hat{\util}_{x}( \PBrule(I^0)).\]
\end{definition}

If a rule $\PBrule$ is based on a utility function $\mu$ then
we write $\PBrule$ satisfies/does not satisfy donation-no-harm
if it satisfies/does not satisfy $\mu$-donation-no-harm.
\ifshort We will use the same shorthand for the other axioms.\fi

\begin{theorem}\label{lem:Discount-fails-no-harm}
  All $\PBrule \in \aggSet$ fail donation-no-harm even if there are no diversity constraints.
\end{theorem}

\iflong
\begin{proof}
  To show the statement, consider the following \pb{} instance~$I$ with three voters~$1,2,3$ and three projects~$p_1,p_2,p_3$.
  The budget~$B$ is $5$, and there are neither donations nor diversity constraints.
  The costs of the projects, and the preferences of the voters are as follows:

\begin{center}
\begin{tabular}{l*{4}{c}}
          & $c(\cdot)$ & $\sat_1$ & $\sat_2$ & $\sat_3$ \\
\midrule                                                
$p_1$     &   $2$      &      $6$ &    $2$   &   $2$     \\
$p_2$     &   $4$      &      $1$ &    $4$   &   $4$     \\
$p_3$     &   $3$      &      $0$ &    $5$   &   $3$     \\
\end{tabular}
\end{center}

There are two feasible and exhaustive bundles $A_1=\{p_1,p_3\}$ and $A_2=\{p_2\}$.
One can verify that~$A_1$ has more score than~$A_2$, under any $\PBrule$.
Hence,~$A_1$ is a winner under~$\PBrule(I)$. 
For instance, $\Sc^{\max}_{\min}(A_1) = 3$ while  $\Sc^{\max}_{\min}(A_2) = 1$.

Now, if voter $3$ donates one unit to~$p_2$, then the reduced cost of $p_2$ will be the same as~$p_3$ and hence, $A_3=\{p_1,p_2\}$ becomes feasible under the new instance.
Indeed, $A_3$ is the unique winner under every $\PBrule \in \aggSet$.
One can verify that this is worse off for voter~$2$ under every~$\utilR$.
For instance, $\Sc^{\max}_{\min}(A_3) = 4$ but~$\util^{\max}_2(A_3)=4 < 5 =\util^{\max}_{2}(A_1)$.
\end{proof}
\fi

\ifshort
\begin{proof}[Proof sketch.]
  To show the statement, consider the following \pb{} instance~$I$ with three voters~$1,2,3$ and three projects~$p_1,p_2,p_3$.
  The budget~$B$ is $5$, and there are neither donations nor diversity constraints.
  The costs of the projects, and the preferences of the voters are as follows:

\begin{center}
\begin{tabular}{l*{4}{c}}
          & $c(\cdot)$ & $\sat_1$ & $\sat_2$ & $\sat_3$ \\
\midrule                                                
$p_1$     &   $2$      &      $6$ &    $2$   &   $2$     \\
$p_2$     &   $4$      &      $1$ &    $4$   &   $4$     \\
$p_3$     &   $3$      &      $0$ &    $5$   &   $3$     \\
\end{tabular}
\end{center}

There are two feasible and exhaustive bundles $A_1=\{p_1,p_3\}$ and $A_2=\{p_2\}$.
One can verify that~$A_1$ is the winner for every $\PBrule \in \aggSet$.

Now, if voter $3$ donates one unit to~$p_2$ then 
$A_3=\{p_1,p_2\}$ becomes feasible.
Indeed, $A_3$ is the unique winner under every $\PBrule \in \aggSet$.
One can verify that this is a worse result for voter~$2$ under $\util^+$ and $\util^{\max}$.
\end{proof}
\fi

\noindent This is a significant drawback as failure to
satisfy donation-no-harm may undermine the acceptance of the 
\pb{} process with donations, especially for voters that cannot 
afford to donate money.
On the other hand, it is easy to see that Pareto-$\PBrule$ and
Sequential-$\PBrule$ satisfy $\hat{\util}$-donation-no-harm by definition.

\begin{proposition}
Pareto-$\PBrule$ satisfies donation-no-harm for all $\PBrule \in \aggSet$.
Sequential-$\PBrule$ satisfies $\hat{\util}$-donation-no-harm
for any monotonic utility function $\hat{\util}$, 
i.e., any utility function $\hat{\util}$ such that $A \subseteq B$ implies $\hat{\util}_i(A) \leq \hat{\util}_i(B)$ \newH{for all voters~$i$.}
\end{proposition}

\iflong
\begin{proof}
Pareto-$\PBrule$: By definition, Pareto-$\PBrule(I^0) = \PBrule(I^0)$ and
Pareto-$\PBrule(I)$ $\hat{\util}$-Pareto-dominates $\PBrule(I^0)$. Hence, by
definition of Pareto-dominance, we have for every voter $i$
that $\hat{\util}_i(\text{Pareto-}(\PBrule,\agg)(I)) \geq \hat{\util}_i(\text{Pareto-}(\PBrule,\agg)(I^0))$.

Sequential-$\PBrule$: By definition,
$\text{Sequential-}\PBrule(I^0) \subseteq \text{Sequential-}\PBrule(I)$. Therefore, by the monotonicity
of $\hat{\util}$ for all voters $i$ we have $\hat{\util}_i(\text{Sequential-}\PBrule(I^0)) \leq \hat{\util}_i(\text{Sequential-}\PBrule(I))$.
\end{proof}
\fi

This motivates us to take a closer look at the different methods for handling donations.
Generally, we want to ensure that it is not harmful to donate more money. However,
it is important to distinguish for whom it should not be harmful if more money is donated.
Here, we can consider three goals: It should not be harmful for a project if more money is donated for that project,
it should not be harmful for the society if more money is donated, and it should not be harmful for a voter if he\todo{One reviewer complained about the incosistency of using ``she'' and ``he'' to refer to a voter. I don't mind.} donates more money.
First, we look at projects:

\begin{definition}[donation-project-monotonicity]\label{def:proj-mono}
  An aggregation rule~$\PBrule$ satisfies \myemph{donation-project-monotonicity} if for each \pb{} instance~$I$,
  each voter~$x$, and each donation~$\contvec'_x$ which is a $j$-variant of $\contvec_x$
  with $\contvec_x[j]<\contvec'_x[j]$  it holds that 
  \[\text{if }j\in \PBrule(I)\text{, then } j\in \PBrule(I-\contvec_x+\contvec'_x).\]
\end{definition}

Increasing the donation for a project~$j$ only makes new bundles available which include~$j$ and
has no effect on the other bundles. Therefore, the following holds:

\begin{proposition}\label{prop:don-proj-mon}
$\PBrule$, Sequential-$\PBrule$ and Pareto-$\PBrule$ satisfy donation-project-monotonicity for all 
$\PBrule \in \aggSet$.
\end{proposition}

\iflong
\begin{proof}
Let $I=(\Pot, \pbinstafter)$ be a \pb{} instance with $\Pot=\profileinstance$,
and $I'=( \Pot-\contvec_x+\contvec'_x, \pbinstafter)$ where
$\contvec'_x$ is a $j$-variant of $\contvec_x$ with $b_x[j]>b'_x[j]$
for a voter~$x$.
Further, let $F = \mathbb{C}(I)$ be the set of feasible bundles in $I$ and $F'=\mathbb{C}(I')$ the set of feasible bundles in $I'$.
We observe that  $F \subseteq F'$ and $j \in C$ for 
all $C \in\mathbb{C}(I') F'\setminus F$.

First, let $\PBrule$ be an aggregation rule based on one of our scoring functions $\agg$.
Then, $\PBrule$ picks the bundle $C$ maximizing $\agg$ among $F$ as the winner 
in $I$. We can distinguish two cases. If there is no $C' \in F' \setminus F$ 
with $\agg(C') \geq \agg(C)$, then $C$ is also the winner in $I'$.
Therefore, donation-project-monotonicity is satisfied. 
On the other hand, if there is a $C' \in F' \setminus F$ such that $\agg(C') \geq \agg(C)$
then the winning bundle $C''$ in $I$ comes from $F'\setminus F$ and hence $j \in C''$.
Hence,  donation-project-monotonicity is also satisfied in this case.

Now, observe that $I^0= (I')^0$. Hence for any arbitrary aggregation rule $\PBrule$
we have $\PBrule(I^0) = \PBrule((I')^0)$. Now, let 
\begin{multline*}
X =\{\PBrule(I^0)\} \cup\\ \{C^* \colon C^* \text{ is feasible and $\util$-Pareto-dominates } \PBrule(I^0)\}
\end{multline*}
and 
\begin{multline*}
X' =\{\PBrule((I')^0)\} \cup \\\{C^* \colon C^* \text{ is feasible and $\util$-Pareto-dominates } \PBrule((I')^0)\}.
\end{multline*}
Then, $X \subseteq X'$ and $j \in C$ for all $C \in X'\setminus X$.
From this, it follows that Pareto-$\PBrule$ satisfies donation-project-monotonicity
as above.

Finally, consider Sequential-$\PBrule$. We observe that for any aggregation rule,
the donation for project $j$ is only considered in the algorithm after $j$ 
is added to the winning bundle or in the final step, which equals just applying $\PBrule$. Hence, Sequential-$\PBrule$ satisfies donation-project-monotonicity if $\PBrule$ satisfies it.
\end{proof}
\fi

Next, we consider the overall satisfaction of the electorate.

\begin{definition}[$\agg$-donation-welfare-monotonicity]\label{def:welfare-mono}
  An aggregation rule $\PBrule$ is \myemph{$\agg$-donation-welfare-monotonicity} if for each \pb{} instance~$I$, %
  each voter~$x$, and
  each contribution vector~$\contvec'_x$ which is a $j$-variant of $\contvec_x$
  with $\contvec_x[j] < \contvec'_x[j]$ %
  it holds that 
  \[\Sc(I, \PBrule(I)) \le \Sc(I',\PBrule(I')) ,\]
  where $I'=I-\contvec_x+\contvec'_x$.
\end{definition}

\iflong
As for $\mu$-donation-no-harm, 
if a rule $\PBrule$ is based on a scoring function $\agg$
we write $\PBrule$ satisfies/does not satisfy donation-welfare-monotonicity
if it satisfies/does not satisfy $\agg$-donation-welfare-monotonicity.
\fi

For each aggregation rule $\PBrule \in \aggSet$, $\PBrule$
and Pareto-$\PBrule$ satisfy 
donation-welfare-monotonicity. This is the case, because increasing the donation only increases the set of bundles over which
$\Sc$ is maximized.
\iflong

\begin{proposition}
For each aggregation rule $\PBrule \in \aggSet$, $\PBrule$
and Pareto-$\PBrule$ satisfy 
donation-welfare-monotonicity.
\end{proposition}

\begin{proof}
Let $I=(\Pot, \pbinstafter)$ be a \pb{} instance with $\Pot=\profileinstance$,
and $I'=( \Pot-\contvec_x+\contvec'_x, \pbinstafter)$ where
$\contvec'_x$ is a $j$-variant of $\contvec_x$ with $\contvec_x[j] < \contvec'_x[j]$.
Then $\mathbb{C}(I) \subseteq \mathbb{C}(I')$. Therefore, we have
\[\max_{A\in \mathbb{C}(I)}\agg_{\aggmode}^{\utilmode}(I,A) \leq \max_{A\in \mathbb{C}(I')}\agg_{\aggmode}^{\utilmode}(I,A)\]
for all $\utilmode \in \{\max,+\}$ and $\aggmode \in \{\sum, \min\}$.
In other words, $\PBrule$ satisfies donation-welfare-monotonicity
for all $\PBrule \in \aggSet$.

Now, consider Pareto-$\PBrule$. For any arbitrary aggregation rule $\PBrule$
we have $\PBrule(I^0) = \PBrule((I')^0)$. Now, let 
\begin{multline*}
X =\{\PBrule(I^0)\} \cup\\ \{C^* \colon C^* \text{ is feasible and $\util$-Pareto-dominates } \PBrule(I^0)\}
\end{multline*}
and 
\begin{multline*}
X' =\{\PBrule((I')^0)\} \cup \\\{C^* \colon C^* \text{ is feasible and $\util$-Pareto-dominates } \PBrule((I')^0)\}.
\end{multline*}
By definition, a change in donation does not change whether one bundle Pareto-dominates
another. Hence $X \subseteq X'$, which implies by the same argument as above that
Pareto-$\PBrule$ satisfies donation-welfare-monotonicity
for all $\PBrule \in \aggSet$.
\end{proof}

\fi
This leaves Sequential-$\PBrule$ %
which does not satisfy donation-welfare-monotonicity.

\begin{proposition}\label{lem:Sequential-fails-welfare-mono}
  Sequential-$\PBrule$ fails donation-welfare-monotonicity for every $\PBrule \in \aggSet$
  even if there are no diversity constraints.
\end{proposition}

\iflong
\begin{proof}
To show the statement, consider the following \pb{} instance with budget $6$:

\begin{center}
\begin{tabular}{l*{8}{c}}
          & $c(\cdot)$ & $\sat_1$ & $\sat_2$ & $\sat_3$ & $b_1$  & $b_2$ & $b_3$ \\
\midrule                                                                 
$p_1$     &   $6$      &      $5$ &    $1$   &   $1$    &  $4$   &   $0$ &  $0$  \\
$p_2$     &   $5$      &      $0$ &    $2$   &   $2$    &  $0$   &   $0$ &  $0$  \\
$p_3$     &   $3$      &      $0$ &    $3$   &   $0$    &  $0$   &   $2$ &  $0$  \\
$p_4$     &   $3$      &      $0$ &    $0$   &   $3$    &  $0$   &   $0$ &  $2$  \\
\end{tabular}
\end{center}

Then, the only feasible, exhaustive bundles in the first iteration of Sequential-$\PBrule$
are either only $p_1$, only $p_2$ or $p_3$ and $p_4$ together.
In each of these bundles, every voter only receives satisfaction from at most 
one project. Hence, it does not matter if $\max$ or $\sum$ is used as utility function.
Now, $p_1$ offers the highest sum of satisfaction and is the only bundle for which the 
minimal satisfaction is not $0$. 
Hence, $p_1$ is chosen in the first iteration of Sequential-$\PBrule$ for any 
$\PBrule \in \aggSet$.

Now, as the donation to $p_1$ is $4$, projects with cost at most $4$ can
be chosen in the second iteration, i.e., only $p_3$ or $p_4$ can be chosen.
In either case, the other project is the only option in a third iteration.
Hence $p_1,p_3,p_4$ are the winning projects for all considered rules.

Now, if voter $1$ increases his donation to $p_1$ by one, then $p_2$ is also a possible solution 
in the second round of the iteration. Indeed, it is chosen by all considered rules
as it has higher sum of satisfaction and higher minimal satisfaction than both $p_3$ and $p_4$.
In this case, there is no third iteration, hence the winning projects are $p_1$ and $p_2$.
It can be checked that this is an overall worse outcome for all considered aggregation methods.
\end{proof}
\fi

The final property asserts that a voter should not be worse off if she decides to 
donate more money to a project. We consider the slightly weaker requirement
that a voter should not be worse off if she donates money for a project 
than if she does not donate any money for that project.
\iflong
Let $\hat{\util}$ denote an arbitrary utility function.
\fi

\begin{definition}[$\hat{\util}$-donation-voter-monotonicity]\label{def:voter-mono}
  An aggregation rule~$\PBrule$ satisfies \myemph{$\hat{\util}$-donation-voter-monotonicity}
  if for each \pb{} instance~$I$, %
  each voter~$x$, and
  each donation~$\contvec'_x$ which is a $j$-variant of $\contvec_x$
  such that $\contvec'_x[j] = 0$ 
   it holds that \[\hat{\util}_{x}(\PBrule(I)) \geq \hat{\util}_{x}( \PBrule(I-\contvec_x+\contvec'_x)).\]
\end{definition}

Unfortunately, this property is essentially impossible to satisfy by an exhaustive rule.
To be more precise, no exhaustive rule which satisfies the following, very weak axiom,
can satisfy $\hat{\util}$-donation-voter-monotonicity, even if we assume satisfaction-consistent contributions.

\begin{definition}\label{def:weak-cont}
An aggregation rule~$\PBrule$ satisfies \myemph{weak continuity}
if for each \pb{} instance~$I$
and for each project $j \in [m]$ the following holds:
If $\sat_i(j) > 0$ for all $i \in [n]$
and there exists a feasible bundle that contains $j$, then
there are values $c$ and $k$ such that $j$ is a winner if
one adds $k$ voters $i_1^*, \dots, i_k^*$ that do not donate any money
and have satisfaction functions such that for all all $\ell \in [k]$ we have~$\sat_{i_\ell^*}(j) = c$
and $\sat_{i_\ell^*}(j^*) =0$ for \newH{all} projects $j^* \neq j$.
\end{definition}

We observe that all our voting rules satisfy weak continuity.%

\begin{lemma}\label{lem:weak-cont}
  $\PBrule$, Sequential-$\PBrule$ and Pareto-$\PBrule$ satisfy weak continuity
for each $\PBrule \in \aggSet$.
\end{lemma}

\iflong
\begin{proof}
Let $I =(\Pot, \pbinstafter)$ be a \pb{} instance.
Let $p \in C$ be a project such that $\sat_i(p) > 0$ for all $i \leq n$.
Now, let $I^c_k$ be the \pb{} instance derived from $I$ by adding
$k$ voters such that $\sat_j(p) = c$
and $\sat_j(q) =0$ for all new voters $j$ and projects $q \neq p$.
By definition, for all new voters $j$, we have for a feasible bundle $C' \in \mathbb{C}(I^c_k)$
and $\util \in \{\max,\sum\}$
\[\util_j(C') = \begin{cases}
c & \text{ if } p \in C',\\
0 & \text{ else.}
\end{cases}\]
Moreover, if $p \in C'$ then, by assumption, $\util_i(C') > 0$ for all voters.
Hence, if $\agg = \min$, then only bundles with $p \in C'$ can be winners if $c > 0$
and $k>0$. Now, let $x \coloneqq \max\{\agg(C')\mid C' \in \mathbb{C}(I)$.
Then, if $c \cdot k > x$, only bundles with $p \in C'$ can be winners
if $\agg = \sum$.
\end{proof}
\fi

\begin{theorem}\label{thm:voter-mono}
No exhaustive aggregation rule which satisfies weak continuity can satisfy
$\utilR$-donation-voter-monotonicity, $\utilmode\in \{\max,+\}$, even
if we only allow satisfaction-consistent contributions
and there are no diversity constraints.%
\end{theorem}

\iflong
\begin{proof}
Consider the following \pb{} instance $I$ with budget $B=4$:

\begin{center}
\begin{tabular}{l*{7}{c}}
          & $c(\cdot)$ & $\sat_1$ & $\sat_2$   &   $b_1$ &   $b_2$ \\
\midrule                                                           
$p_1$     &   $3$     &      $1$ &    $1$     &    $0$  &    $0$  \\
$p_2$     &   $4$     &      $2$ &    $3$     &    $1$  &    $2$  \\
$p_3$     &   $4$     &      $3$ &    $2$     &    $2$  &    $1$  \\
\end{tabular}
\end{center}

We assume that there are $k$ voters such that $\sat_j(p) = c$
and $\sat_j(q) =0$ for all new voters $j$ and projects $q \neq p$
where $k$ and $c$ are such that $p_1$ is in the winner set of 
all instances occurring in this proof.

We observe that the only feasible, exhaustive bundles that contain $p_1$ are
$A_1=\{p_1,p_2\}$ and $A=\{p_1,p_3\}$. Assume first that $A_1$ is the winning bundle. 
Then voter~$3$ can decrease her donation for~$p_2$ to $0$. 
which makes $A_2$ the only feasible, exhaustive bundle containing $p_1$.
Hence $A_2$ must be the winning bundle for any exhaustive rule that satisfies
weak continuity. Now, observe that $\utilR_1(\{p_1, p_3\}) > \utilR_1(\{p_1,p_2\})$ for each $\utilmode \in \{+,\max\}$,
which contradicts $\utilR$-donation-voter-monotonicity.

If $A_2$ is the winning bundle, then since the roles of voter~$2$ and $3$ are symmetric,
voter~$2$ can increase his utility by reducing his donation for~$p_3$ to $0$ by a analogous argument. A contradiction.
\end{proof}
\fi

Observe that for all variations of $\PBrule_{\aggmode}^{\max}$
if we additionally assume that a voter only donates to the projects which give her the 
highest satisfaction,
then donation-voter-monotonicity rules exist.
In particular, this means that for dichotomous 
preferences any version of $\PBrule_{\aggmode}^{\max}$ satisfies donation-voter-monotonicity if we
assume satisfaction-consistent contributions. %

\iflong
\begin{proposition}
$\PBrule_\Sigma^{\max}$ and $\PBrule_{\min}^{\max}$ satisfy donation-voter-monotonicity
if all voters only donate for the projects that gives them the 
highest satisfaction.
\end{proposition}
\begin{proof}
Let $\PBrule$ be either $\PBrule_\Sigma^{\max}$ or $\PBrule_{\min}^{\max}$.
Further, let $I=(\Pot, \pbinstafter)$ be a \pb{} instance with $\Pot=\profileinstance$,
and $I'=( \Pot-\contvec_x+\contvec'_x, \pbinstafter)$ where
$\contvec'_x$ is a $j$-variant of $\contvec_x$ with $b_x[j]>b'_x[j]$
for a voter~$x$.

Then, it follows from the proof of Proposition~\ref{prop:don-proj-mon} that 
$\PBrule(I) \neq \PBrule(I')$ implies $j \in \PBrule(I')$. Therefore,
$\util_x^{\max}(\PBrule(I')) = \sat_i(j)$. As $x$ donates money for $j$,
our assumption implies that $\sat_i(j) \geq \sat_i(j')$ for all $j' \in [m]$.
Therefore, $\util_x^{\max}(\PBrule(I')) \geq \util_x^{\max}(C')$ for all 
$F \in \mathbb{C}(I)$ -- including $\PBrule(I)$ -- by the definition of $\util^{max}_i$.
\end{proof}
\fi 

\section{Central Computational Problems}

\looseness=-1
We consider two decision problems that arise in our framework.
The first one captures the complexity of applying an aggregation method.

\decprob{\probWinner}
{A \pb{} \iflong instance~$I=(\Pot, \pbinstafter)$ with $\Pot=\profileinstance$, and a bundle~$A\subseteq [m]$.
  \fi
}
{Is $A$ a \emph{(co-)winning bundle} under~$\PBrule$?}

\noindent Note that if the aggregation rule~$\PBrule$ is defined based on a scoring function~$\Sc$, then to decide whether $A$ is a (co)-winning bundle under~$\PBrule$ means to decide whether $\Sc(I,A) \ge \Sc(I,A')$ holds for all feasible bundles~$A'\in \mathbb{C}(I)$. 

\iflong
To show hardness of \probWinner, we will consider its co-problem, which is defined as follows:

\decprob{\probNotWinner}
{A \pb{} instance~$I=( P, \pbinstafter)$ with $P=\profileinstance$, and a bundle~$A\subseteq [m]$.}
{Is $A$ \emph{not} a (co-)winning bundle under~$\PBrule$?}

Note that if the aggregation rule~$\PBrule$ is defined based on a scoring function~$\Sc$,
then to decide whether $A$ is not a (co)-winning bundle means to decide whether there exists a feasible bundle~$C'\in \mathbb{C}(I)$ such that $\Sc(C') > \Sc(A)$ holds.
For such aggregation rule, we say that \myemph{a bundle~$X$ defeats another bundle~$Y$ under $\PBrule$} if $\Sc(I,X) > \Sc(I,Y)$.
\fi

The second problem is concerned with the effective use of donations from a voter's perspective.
This problem is particularly crucial in light of Theorem~\ref{thm:voter-mono}, which tells us that
voters need to carefully consider how to distribute their donation independently of the voting rule used.
\iflong
\decprob{\probSpend}
{%
  A \pb{} instance~$I=( \Pot, \pbinstafter)$ with $\Pot=\profileinstance$, a voter~$v\in [n]$, and a donation bound~$\don\in \mathds{N}$.%
}
{Is there a donation vector~$\contvec'_v\in \mathds{N}^m$  with $\sum_{j\in [m]}{\contvec'_v} \le \don$ such that $\util_v(\PBrule(I')) > \util_v(\PBrule(I))$,
  where $I'\coloneqq I-\contvec_v+\contvec'_v$ and $\util$ denotes the utility function associating with the aggregation rule~$\PBrule$?}

  \fi
  
\iflong Before we turn to the algorithmic complexity, we discuss some structural property. 

\begin{lemma}\label{lem:prem-max-+}
  For each two bundles~$X,Y\subseteq C$ with $|Y|=1$ it holds that
  if $X$ defeats $Y$ under~$\PBrule^{\max}_{\min}$, then $X$ defeats $Y$ under~$\PBrule^{+}_{\min}$.
\end{lemma}

\begin{proof}
  Let $I=( P, \pbinstafter)$ be a \pb{} instance with $P=\profileinstance$.
  Further, let $X$ and $Y$ be two bundles of $C$ with $|Y|=1$.
  Assume that $X$ defeats $Y$ under~$\PBrule^{\max}_{\min}$, meaning that  $\Sc^{\max}_{\min}(X) > \Sc^{\max}_{\min}(Y)$.
  By definition, this implies that
  \begin{align*}
    \min_{i\in [n]} \max_{j\in X}\sat_{i}(j) > \min_{i\in [n]}\max_{j \in Y}\sat_i(y) \stackrel{|Y|=1}{=} \min_{i \in [n]}\sat_i(Y).
  \end{align*}
  Since %
  $\min\limits_{i\in [n]}\sat_{i}(X)\ge \min\limits_{i\in [n]}\max\limits_{j\in X}\sat_{i}(j)$,
  it follows that
  $\Sc^{+}_{\min}(X)=\min\limits_{i\in [n]}\sat_{i}(X) > \min\limits_{i\in [n]}\sat_{i}(Y)=\Sc^{+}_{\min}(Y)$, i.e., $X$ defeats $Y$ under~$\PBrule^{+}_{\min}$, as desired.
\end{proof}
\fi

\subsection{\probWinner}

\iflong \begin{table*}
  \caption{Complexity results of \probWinner. Here, $0/1$ means dichotomous preferences. All hardness results hold even if there are no donations.}\label{table:results-winner}
  \centering
  \begin{tabular}{@{}l|l@{\;}l@{\;}l@{\qquad}l@{\;}l@{\;}l@{}}
    \toprule
    Utility~$\util^{\utilmode}$ & \multicolumn{3}{c}{$\util^{\max}$} & \multicolumn{3}{c}{$\util^{+}$} \\\midrule
    $\PBrule^{\utilmode}_{\min}$ & coNP-hard & ($t=0$, $0/1$, unit cost) &  \mythm{thm:winner-coNPh-(max,min)-(+,min)-(max,sum)} & coNP-hard & ($t=0$, $0/1$, unit cost) & \mythm{thm:winner-coNPh-(max,min)-(+,min)-(max,sum)} \\
    $\PBrule^{\utilmode}_{\Sigma}$ & coNP-hard & ($t=0$, $0/1$) & \mythm{thm:winner-coNPh-(max,min)-(+,min)-(max,sum)} & coNP-hard & ($t$ unbounded, $0/1$) & \mythm{thm:coNP-hard-+sum}  \\
    \bottomrule
\end{tabular}
\end{table*}
\fi

In this section, we investigate the algorithmic complexity of \probWinner \iflong (also see \cref{table:results-winner} for an overview)\fi.{}
First of all, we establish the complexity upper bound.

\begin{theorem}
  \probWinnerR{$\PBrule$} is in coNP for each $\PBrule \in \aggSet$.
\end{theorem}

\iflong
\begin{proof}
  Let~$I=(\Pot, \pbinstafter)$ with $\Pot=\profileinstance$ be a \pb{} instance, and let $A$ be given bundle.
  It suffices to show that \probNotWinner\ is in NP.
  If $A$ is not a winner, i.e., $(I,A)\in$ \probNotWinnerR{$\RR$}, then there must be another feasible bundle~$C'\in \mathbb{C}(I)$ which has a higher score than~$A$, i.e.,
  \begin{align}\label{eq:notWinner}
    \Sc(I,C') > \Sc(I,A).
  \end{align}
  Since given a second feasible bundle~$C'\in \FB$, in polynomial time, we can compute the utility function~$\Sc$ used under~$\RR$ and check the inequality given in~\eqref{eq:notWinner}, we have a polynomial certificate for~$(I,A)$ regarding \probNotWinner.
  This immediately implies that \probNotWinner and \probWinner are in NP and coNP, respectively.
\end{proof}
\fi
\iflong

\subsubsection{The \probWinnerR{$\PBrule^{+}_{\Sigma}$} problem}

We start with $\PBrule^{+}_{\Sigma}$ rule and observe that the complexity of determining a winner depends on the number of types.
\fi

\ifshort
\newH{Since all aggregation rules considered in this paper generalize commonly used multiwinner voting rules, it is fairly straight-forward to obtain the following hardness results.}
\fi

\begin{theorem}\label{thm:coNP-hard-+sum}
  \probWinnerR{$\PBrule^{+}_{\Sigma}$} can be solved in $O(\dptime)$~time.
  It is coNP-hard even for dichotomous preferences.
  It is weakly coNP-hard even if there are no diversity constraints.
\end{theorem}

\iflong
\begin{proof}
  \mypa{The first statement.}
  To show the running time, we provide a straight-forward dynamic program~(DP) similar to the one for the weakly NP-hard \knapsack problem.
  To this end, let $(I=( P, \pbinstafter), A)$ denote an instance of \probWinnerR{$\PBrule^{+}_{\Sigma}$} with $P=\profileinstance$.
  Further, for each project~$j\in [m]$, define the reduced cost~\myemph{$\cost^*_j \coloneqq \max(\cost_j-\sum_{i\in [n]}\contvec_i[j], 0)$} and the additive score~\myemph{$\score_j\coloneqq \sum_{i\in [n]}\sat_{i}(j)$}.

  The idea behind the DP approach is to create a table, which stores, for each configuration (including budget) and each~$j\in [m]$, the highest sum of satisfaction of bundles satisfying that specific configuration.
  More precisely, for each budget~$B' \le B$, each vector~$\boldsymbol{\tau}\in [m]^{t}$,
  and each project~$j\in [m]$,
  we aim to compute and store the value $S(B', \boldsymbol{\tau}, j)$, which
  is the maximum of the sum of the satisfactions of voters towards all bundles~$C'\subseteq [j]$ which have cost at most~$B'$ and meets the diversity constraints defined by $\boldsymbol{\tau}$.
  Formally,
  \begin{align}
    S(B', \boldsymbol{\tau}, \boldsymbol{f}, j) = \max_{\substack{C'\subseteq [j] \text{ s.t. }\\
    \sum\limits_{x\in C'}\cost^*_x  \le B',
    \\ \sum\limits_{x\in C'}\typevec_x = \boldsymbol{\tau}}}\sum_{j\in C'} \score_j.
  \end{align}

\noindent  We initialize and build the table as follows:
  \begin{align*}
 &    S(0, 0, 0)  \coloneqq  0\\
  &  S(B',\boldsymbol{\tau}, j)\\
   &  \coloneqq
      \begin{cases}
        S(B',\boldsymbol{\tau},  j-1),~ ~~~ \text{if } B' < \cost^*_j \text{ or } \exists z\in [t]\colon   \boldsymbol{\tau}[z] < \typevec_j[z]\\
        \max\{S(B',\boldsymbol{\tau}, j-1),\\
        \qquad~ S\big(B'-\cost^*_j,
        \boldsymbol{\tau}-\typevec_j, j-1\big)+\score_j\},  \text{otherwise.}
  \end{cases} 
  \end{align*}

  To decide~$(I,A)$ we check whether for each type configuration~$\boldsymbol{\tau}\in [m]^{t}$ with $\lowervec\le \boldsymbol{\tau} \le \uppervec$  
  it holds that $\aggsum^{+}(I,A) \ge S(B,\boldsymbol{\tau},m)$.
  We answer yes if the above holds, otherwise we answer no.

  The running time of the DP approach depends on the running time of computing the table.
  Note that if we pre-compute for each project~$j\in [m]$ its gain of sum of satisfactions~$\score_j$ and its reduced cost~$\cost^*_j$,
  then we can complete the operation ``maximum'' for each entry in the above table in $O(t)$~time.
  This preprocessing runs in $O(n\cdot m)$ time.
  Since the table has $O((B+1)\cdot (m+1)^{t+1})$ entries, each of which can be computed in $O(t)$~time, in total the algorithm takes $O(\dptime)$ time. %

  \mypa{The coNP-hardness statement.}  As for the hardness result,
  we reduce from the NP-complete \textsc{Vertex Cover} problem, defined as follows:

  \decprob{Vertex Cover}{
    A graph~$G=(U,E)$ and a non-negative integer~$k$}
  {Does $G$ admit a \myemph{vertex cover} of size at most~$k$, i.e., a size-at-most-$k$ vertex subset~$U'\subseteq U$ such that for each edge~$e_i \in E$ it holds that $e_i\cap U'\neq \emptyset$?}
  We will introduce unbounded number of types, as the problem is pseudo-polynomial for constant number of types (indicated by the dynamic program above).
  The constraints on these types will ensure that the sought vertex subset is a vertex cover.

  Let $(G=(U,E), k)$ be an instance of \textsc{Vertex Cover} with $U=\{u_1,u_2,\ldots, u_{\enn}\}$ and $E=\{e_1,e_2,\ldots, e_{\emm}\}$.
  Without loss of generality, let us assume that $G$ contains two disjoint edges such that any vertex cover of $G$ has size at least two.
  Again, to show coNP-hardness of \probWinnerR{$\PBrule^{+}_{\Sigma}$}, we show NP-hardness of \probNotWinnerR{$\PBrule^{+}_{\Sigma}$}.
  The intended instance of \probNotWinnerR{$\PBrule^{+}_{\Sigma}$} consists of $\enn+1$ projects~$u_0,u_1,\ldots,u_{\enn}$, one voter~$w$, and $\emm$~types, i.e., $t = \emm$.
  Thus, the set of projects is defined as~$C\coloneqq \{u_0,u_1,\ldots,u_{\enn}\}$.
  The satisfaction of voter~$w$ towards each project from~$C$ is one.
  The cost of each project, except project~$u_0$, is one, while the cost of project~$u_0$ is~$k$.
  The projects have the following types:
  \begin{align*}
    \typevec_{u_0}  & =  1^{\emm}\\
    \typevec_{u_i}[z] & =
                        \begin{cases}
     1, & \text{ if } u_i \in e_z\\
     0, & \text{ otherwise.}
    \end{cases}
  \end{align*}
  In other words, the types encode the incidence matrix of the graph.
  Let the lower bound and the upper bound of the diversity constraints be $1^{\emm}$ and $2^{\emm}$, respectively.
  Set the budget~$B\coloneqq k$.
  The target bundle~$A$ consists of only one project, namely~$u_0$.
  This completes the description of the construction, which can clearly be done in polynomial time.
  Let $I$ denote the constructed \pb{} instance.
  
  It remains to show the correctness, i.e., $G=(U,E)$ has a vertex cover of size at most~$k$ if and only if
  $A=\{u_0\}$ is \emph{not} a winner under~$\PBrule^{+}_{\Sigma}$.
  For the ``only if'' part, assume that $G=(U,E)$ admits a vertex cover~$U'$ of size~$k$.
  We claim that $A$ is not a winner under~$\PBrule^{+}_{\Sigma}$, by showing that $U'$ is a feasible bundle and defeats $A$ under~$\PBrule^{+}_{\Sigma}$.

  We first show that $U'$ is feasible.
  Clearly, $U'$ costs at most~$k$, and hence the budget constraint is satisfied.
  As for the diversity constraint, since $U'$ is a vertex cover, for each type~$z\in [\emm]$,
  there exists at least one project~$u_i\in U'$ with $\typevec_{u_i}[z]=1$.
  Furthermore, by definition, we cannot have $\typevec_{u_i}[z]>2$

  To show that $U'$ defeats $A$, recall that any vertex cover of $G$ contains at least two vertices.
  Hence, $\Sc^{+}_{\Sigma}(I, U') = \sat_{w}(U') \ge 2 > 1 = \sat_{w}(A) = \Sc^{+}_{\Sigma}(I, A)$.

  For the ``if'' direction, assume that $A$ is not a winner under~$\PBrule^{+}_{\Sigma}$.
  This means that there exists a feasible bundle~$C'\subseteq C$ which has higher score than~$A$, i.e.,
  $\sat_{w}(C') > \sat_{w}(A)=1$.
  For ease of notation, define the vertex subset~$U'\coloneqq C'\setminus \{u_0\}$; recall that $C=U\uplus \{u_0\}$.
  We claim that $U'$ is a vertex cover of size at most~$k$.
  It is straight-forward to see that $|U'|\le k$ since each project in~$C\cap U$ has unit cost.
  To see that $U'$ is indeed a vertex cover, suppose, for the sake of contradiction, that there exists an edge~$e_j\in E$ with $U'\cap e_j =\emptyset$.
  By the definition of types, this means that $\sum_{u_i\in U'}\typevec_{u_i}[j]=0$.
  However, by the diversity constraint that $\lowervec[j] = 1$, it follows that $u_0\in C'$ since besides the projects which correspond to the endpoints of~$e_j$,
  project~$u_0$ is the only project which has type~$j$.
  However, this would imply that $C'=\{u_0\} = A$ since the cost of~$u_0$ is $k$ and it is not possible to add any other project to~$C'$ because of the budget constraint.
  This is a contradiction to $C'$ be a defeater of~$A$.

  \mypa{The weakly coNP-hardness statement.}   It is quite straight-forward to see that
  the weakly NP-complete \knapsack problem can be reduced to \probNotWinnerR{$\PBrule^{+}_{\Sigma}$} with $t=0$:
  Create a \myemph{item-project} for each item with cost equals to the weight of the item,
  and a voter~$v$ whose satisfaction for each item-project corresponds to the value of the item;
  create an additional project~$\alpha$ with cost~$S$ (the weight bound) for which voter~$v$ has satisfaction exactly one less than the given value lower bound of the \knapsack instance.
  One can verify that there exists a subset of items with weight at most~$S$ and value at least~$K$
  iff.
  the bundle~$\{\alpha_0\}$ is not a winner.
\end{proof}
\fi

\iflong
\subsubsection{The remaining \probWinner{} problems}

In this subsection, we consider the algorithmic complexity of deciding a winner when the aggregation is of egalitarian nature: $\min$, or when the utility function uses $\max$. 

\fi
\begin{theorem}\label{thm:winner-coNPh-(max,min)-(+,min)-(max,sum)}
  For each $\PBrule \in \{\PBrule^{\max}_{\min}, \PBrule^{+}_{\min}, \PBrule^{\max}_{\Sigma}\}$, \probWinnerR{$\PBrule$} is coNP-hard even if the projects have unit costs, and there are neither diversity constraints, nor donations.
  Hardness for $\PBrule^{\max}_{\min}$ and $\PBrule^{+}_{\min}$ remains even if the voters have dichotomous preferences.
\end{theorem}

\iflong
\begin{proof}
  As already mentioned, to show the coNP-hardness of \probWinner{}, we show NP-hardness of its co-variant~\probNotWinner{}, by reducing from the NP-complete \textsc{Vertex Cover} problem~(see the proof of \cref{thm:coNP-hard-+sum} for the formal definition).
  
  Let $(G=(U,E), k)$ denote an instance of \textsc{Vertex Cover} with vertex set~$U=\{u_1,u_2,\ldots, u_{\enn}\}$ and edge set~$E=\{e_1,e_2,\ldots,e_{\emm}\}$.

  We first consider the rules~$\PBrule^{\max}_{\min}$ and $\PBrule^{+}_{\min}$, and then the rule~$\PBrule^{\max}_{\Sigma}$.

  \myparagraph{The rules~$\PBrule^{\max}_{\min}$ and $\PBrule^{+}_{\min}$.}
  We construct the same \pb{} instance for both hardness reductions and we will prove the correctness of the construction separately.

  The \pb{} profile~$P$ in the intended \pb{} instance consists of exactly $\enn+1$ projects, called~$u_0,u_1,\ldots,u_{\enn}$, and $\emm$~voters, called~$e_1,e_2,\ldots,e_{\emm}$.
  Note that, to connect the two problems, we use the same notation for the vertices (resp.\ edges) and for the projects (resp.\ voters).
  The meaning of the notations will be clear from the context.

  For the sake of brevity, let $C\coloneqq \{u_0,u_1,\ldots,u_{\enn}\}$ %
  denote the set of projects. %
  The cost~$\cost_{u_i}$ of each project~${u_i}\in C$ is one and the budget is set to~$B\coloneqq k$.
  The bundle~$A$ that we are interested in consists of only one project, namely~$u_0$, i.e., $A\coloneqq \{u_0\}$.
  
  The satisfaction of each voter~$e_j\in E$, $j\in [\emm]$, towards each project~$u_i$, $i\in \{0,1,\ldots,\enn\}$, is defined as follows:
  \begin{align*}
    \sat_{e_j}(u_i)\coloneqq
    \begin{cases}
       1, & \text{ if } i\neq 0 \text{ and } u_i \in e_j,\\
       0, & \text{ otherwise }.
    \end{cases}
  \end{align*}
  Briefly put, the satisfaction models the incidence matrix of the graph~$G$.

  We do not impose any restriction on the diversity constraints and no voter donate any money.
  For the sake of completeness, for each voter~$e_j\in E$ define the donation~$\contvec_{e_j} \coloneqq \zeros$,
  and let $t=0$. %
  Then, the constructed instance is defined as follows~$I=(\Pot, B=k,\emptyset, \emptyset)$ with $\Pot=(t=0, (\cost_{u_i})_{u_i\in C}, (\typevec_{u_i})_{u_i\in C}, (\sat_{e_j})_{e_j\in E}, (\contvec_{e_j})_{e_j\in E})$.

  This completes the description of the instance, which, clearly, can be constructed in polynomial time.
  One can also verify the required conditions given in the statement are also satisfied.

  It remains to show the correctness.
  Indeed, we show that
  \begin{compactenum}[(i)]
    \item\label{st:VC->min-max+} If $(G,k)$ is a yes-instance of \textsc{Vertex Cover}, then $(I, A)$ is a yes-instance of \probNotWinnerR{$\PBrule^{\max}_{\min}$} and \probNotWinnerR{$\PBrule^{+}_{\min}$}, i.e., $A$ is \emph{not} a winner under~$\PBrule^{\max}_{\min}$ or~$\PBrule^{\max}_{\min}$.
    \item\label{st:min-max+->VC} If $(I, A)$ is a yes-instance of \probNotWinnerR{$\PBrule^{\max}_{\min}$} or \probNotWinnerR{$\PBrule^{+}_{\min}$},
      then
      $(G,k)$ is a yes-instance of \textsc{Vertex Cover}.
  \end{compactenum}

  For Statement~\eqref{st:VC->min-max+}, assume that $(G,k)$ is a yes instance and let $U'\subseteq U$ denote a size-at-most-$k$ vertex cover of~$G$.
  We first show that $U'$ defeats $A$ under~$\PBrule^{\max}_{\min}$; this suffices to show that $A$ is not a winner under~$\PBrule^{\max}_{\min}$ since $|U'|\le k$, meaning that $U'$ is a feasible bundle.
  Since $U'$ is a vertex cover, for each edge~$e_j\in E$ we have that $e_j\cap U' \neq \emptyset$.
  By our construction of the satisfaction, this implies that $\Sc^{\max}_{\min}(U') = \min_{e_j\in E}\max_{u_i \in U'} = 1$.
  Since for each voter~$e_j\in E$ it holds that $\sat_{e_j}(A)  = 0 = \sat_{e_j}(\{u_0\})$, and hence $\Sc^{\max}_{\min}(A)=0$, it follows that $U'$ defeats~$A$ under~$\PBrule^{\max}_{\min}$.
  That $U'$ defeats~$A$ also under~$\PBrule^{+}_{\min}$ follows from~\cref{lem:prem-max-+} (applied for $X=U'$ and $Y=A$).

  For Statement~\eqref{st:min-max+->VC}, we show the contra-positive, i.e., if $(G,k)$ is a no-instance of \textsc{Vertex Cover}, then $A$ is a (co-)winner under~$\PBrule^{\max}_{\min}$ and $\PBrule^{+}_{\min}$.
  Assume that $(G,k)$ is a no-instance, i.e., for each size-$k$ vertex subset~$U'\subseteq U$, there exists an edge~$e_j\in E$ with $U'\cap e_j = \emptyset$.
  Now, consider an arbitrary feasible bundle, i.e., a subset of projects~$C'\subseteq C$ with $|C'|\le B = k$.
  For ease of notation, define the vertex subset~$U'\coloneqq C'\setminus \{u_0\}$; recall that $C=U\uplus \{u_0\}$.
  Clearly, $|U'|\le k$, and by assumption, there exists an edge~$e_j\in E$ with $U'\cap e_j = \emptyset$.
  Since $U'$ is also a subset of projects, by the definition of satisfaction, for this voter~$e_j$, it holds that
  $\sat_{e_j}(C') \le \sat_{e_j}(u_0) + \sat_{e_j}(U') = 0$.
  This immediately imply that $\min\limits_{e \in E}\max\limits_{x\in C'}\sat_{e}(x) \le \min\limits_{e \in E}\sat_{e}(C')=0 = \min\limits_{e\in E}\sat_{e}(A)$.
  Hence, $A$ is not defeated by any feasible bundle, neither under~$\PBrule^{\max}_{\min}$ nor under $\PBrule^{+}_{\min}$.

  \myparagraph{The rule~$\PBrule^{\max}_{\Sigma}$.}
  The construction is similar to the one above.
  The major difference is that due to the nature of the utility function~$\util^{\max}$,
  we do not need any type to show hardness.

  We construct a \pb{} instance~$I$ as follows.
  It consists of zero types, $\enn+1$~projects, called~$u_0,u_1,\ldots,u_{\enn}$, and $2\emm-1$~voters, called~$e_1,e_2,\ldots,e_{\emm}$, and $f_1,f_2,\ldots, f_{\emm-1}$.
  Thus, the set of projects is defined as~$C\coloneqq U\cup \{u_0\}$.
  To ease notation, define $F\coloneqq \{f_1,f_2,\ldots,f_{\emm-1}\}$.
  Each project, except~$u_0$, has a unit cost, while project~$u_0$ has cost~$k$.
  The budget~$B$ is set to exactly~$k$.
  Each voter~$e_j$, $j\in [\emm]$,  has a satisfaction of value~$1$ to a project~$u_i$ if and only if $u_i\in e_j$.
  \begin{align*}
    \sat_{e_j}(u_i)\coloneqq
    \begin{cases}
       1, & \text{ if } i\neq 0 \text{ and } u_i \in e_j,\\
      0, & \text{ otherwise }.
    \end{cases}
  \end{align*}
  Briefly put, the satisfactions of the voters from~$E$ model the incidence matrix of the graph~$G$.
  Each voter~$f_j \in F$ has satisfaction one to project~$u_0$ and zero to the remaining projects.
  That is, $\sat_{f_j}(u_i) = 0$ for all $u_i\in U$, and $\sat_{f_j}(u_0)=1$.

  We do not impose any restriction on the diversity constraints and no voter donates any money.
  For the sake of completeness, for each voter~$e_j\in E$ define the donation~$\contvec_{e_j} \coloneqq \zeros$,
  and define the diversity constraints~$\lowervec=\uppervec\coloneqq \emptyset$.
  Summarizing, the constructed instance is~$I=(\Pot, B=k, \emptyset,\emptyset)$ with  $\Pot=(t=0, (\cost_{u_i})_{u_i\in U}, 0, (\sat_{e})_{e \in E\cup F}, \zeros)$.
  The initial bundle~$A$ consists of only project~$u_0$.
  This completes the description of the instance, which can clearly be done in polynomial time.
  
  It remains to show the correctness, i.e., $G=(U,E)$ admits a vertex cover of size at most~$k$ if and only if $A$ is not a winner under~$\PBrule^{\max}_{\Sigma}$.
  For the ``only if'' direction, assume that $U'\subseteq U$ is a vertex cover of size at most~$k$.
  Then, it is straight-forward to verify that $U'$ is a feasible bundle and $\sum_{e \in E\cup F} \sat_{e}(U') = m > m-1 \sum_{e\in E\cup F}\sat_e(A) = m-1$.
  This means that $U'$ is a feasible defeater of~$A$.

  For the ``if'' direction, assume that $U'\subseteq C$ is a feasible bundle and it defeats~$A$.
  We first observe that $u_0\notin U'$ as otherwise, by the budget constraint, it must hold that $U'=\{u_0\}$ since the cost of~$u_0$ is~$k$.
  Hence, $U'$ cannot defeat~$A$.  
  This implies that $U'\subseteq U$.
  Now, we claim that $U'$ is a vertex cover of size at most~$k$.
  By to the budget constraint and by the unit costs of the projects in $U$, it follows that
  $|U'|\le k$.
  
  It remains to show that $U'$ is a vertex cover.
  Suppose, for the sake of contradiction, that $e_j\in E$ is an edge which is not ``covered'' by~$U'$, i.e.,
  $U'\cap e_j=\emptyset$.
  Then, for the corresponding voter~$e_j$, it must hold that $\sat_{e_j}(U')=0$.
  Consequently, we deduce that
  \begin{align*}
    \Sc^{\max}_{\Sigma}(I, U') & = \sum_{w\in E\cup F}\max_{u_i\in U'}\sat_{w}(u_i) \\
                             & = \sum_{w\in E\setminus \{e_j\}}\max_{u_i\in U'}\sat_{w}(u_i)\\
                               & \le m-1 = \Sc^{\max}_{\Sigma}(I,A), 
  \end{align*}
  a contradiction to $U'$ having defeated~$A$.
\end{proof}
\fi

We observe that $\PBrule$, Sequential-$\PBrule$ and Pareto-$\PBrule$\footnote{We 
say a bundle $A$ is a co-winner under Pareto-$\PBrule$ if there is a tie-breaking
order under which $A$ is a winner.} coincide if there are no donations.
Since all hardness reductions above do not require any donations, they can be used to directly show the following.
\begin{corollary}\label{cor:winner-seq}
  All hardness results stated in Theorems~\ref{thm:coNP-hard-+sum}--\ref{thm:winner-coNPh-(max,min)-(+,min)-(max,sum)} hold for the corresponding Sequential- and Pareto-variants.
\end{corollary}

\subsection{\probSpend}
In this section, we investigate the complexity of finding an effective donation.
We prove that when there are no diversity constraints, for all aggregation rules~$\RR$ except $\PBrule^{+}_{\Sigma}$, finding an optimal donation is as hard as the complexity class~\parallelnp~\cite[Chapter 17.1]{Pap94}, which includes~NP.
For $\PBrule^{+}_{\Sigma}$, it is both weakly NP-hard and weakly coNP-hard.
This implies, under a widely believed complexity-theoretical assumption, that our problem is beyond NP.
Moreover, if diversity constraints are present, \probSpend{} is even \sigmatwop-complete for all aggregation rules and their two variants considered in this paper.

\iflong
\myparagraph{Remark.}
  \parallelnp (or P$^{\text{NP}[\log]}$, aka.~$\Theta^{\text{P}}_2$~\cite{Wagner1990}) contains all problems solvable by some P oracle machine that, instead of asking its oracle queries sequentially, accesses its NP oracle in parallel.
  There are well-known \parallelnp-complete voting problems such as determining whether a given candidate is a winner under the Dodgson, Kemeny, or Young voting rule~\cite{HHR97,HSV05,RSV03}.
  \sigmatwop{} is a complexity class from the second level of the polynomial hierarchy,
and problems known to be complete for this class include deciding core stability in additive hedonic games~\cite{Woeginger2013}, stable matching with diversity constraints~\cite{ChenGanianHamm2020ijcai-diversestable}. %
\fi
We recall the following relations among the complexity classes:
\iflong
\begin{align*}
  (\text{NP} \cup \text{coNP}) \subseteq \text{\parallelnp} \subseteq \text{\sigmatwop},
\end{align*}
\fi

\noindent where all inclusions are generally assumed to be strict.

In the following, after locating the complexity upper bound, we first consider the case without diversity constraints, and then that with diversity constraints.

\begin{theorem}\label{thm:spend-in-sigma2p}
   \probSpendR{$\PBrule$}, \probSpendR{\text{\normalfont Sequential-}$\PBrule$}, and  \probSpendR{\text{\normalfont Pareto-}$\PBrule$} are in \sigmatwop{} for each $\PBrule \in \aggSet$.
\end{theorem}

\iflong
\begin{proof}[Proof sketch.]
  To prove that \probSpendR{$\RR$} is in \sigmatwop, it suffice to show that the problem can be reduce to a problem known to be in \sigmatwop, namely, \twoqsat.
  The \twoqsat problem is defined as follows:
  
  \decprob{\twoqsat}
  {Two sets~$X$ and $Y$ of Boolean variables; a Boolean formula~$\phi(X,Y)$ over~$X\cup Y$.}
  {Does there exist a truth assignment of $X$ such that for each truth assignment of $Y$ the formula~$\phi(X,Y)$ is evaluated to true?}

  Let $\utilmode\in \{\max,+\}$ and $\aggmode \in \{\Sigma, \min\}$.
  Further let $(I, v, \don)$ be an instance of \probSpendR{$\RR$} with $I=(\Pot, \pbinstafter)$ and $\Pot=\profileinstance$.
  Then, to decide \probSpendR{$\RR$}, we can equivalently ask whether there exists a contribution vector~$\contvec'_v\in \mathds{N}^m$ and two ``winning'' bundles~$A_0, A_1 \subseteq C$ (with respect to two different instances) such that
  for each two bundles~$A_2,A_3\subseteq C$ the following holds:
  \begin{compactenum}[(i)]
    \item $\sum_{j\in [m]}\contvec'_v[j] \le \don$.
    \item $\util_{v}^{\utilmode}(A_1) > \util_{v}^{\utilmode}(A_0)$.
    \item $A_1$ is a winning bundle for $I'=I-\contvec_v+\contvec'_v$, i.e.,
     $A_1$ is feasible~$I'$, and if $A_3$ is feasible for $I'$, then $\ScR(A_1)\ge \ScR(A_3)$.
    \item $A_0$ is a winning bundle for $I$, i.e.,
       $A_0$ is feasible for~$I$, and if $A_2$ is feasible for $I$, then $\ScR(A_0)\ge \Sc(A_2)$.
  \end{compactenum}
  It is straight-forward to verify that the above conditions can be encoded via an \twoqsat{} instance.

  \newH{  The Pareto-variant of~$\RR$ can also be encoded using a similar approach.
    We ask whether there exists a contribution vector~$\contvec'_v\in \mathds{N}^m$ and one initial winner~$C_0\subseteq C$ and two ``Pareto-optimal'' bundles~$A_0, A_1 \subseteq C$ (with respect to two different instances) such that
  for each two bundles~$C'\subseteq C$ the following holds:
  \begin{compactenum}[(i)]
    \item $\sum_{j\in [m]}\contvec'_v[j] \le \don$.
    \item $\util_{v}^{\utilmode}(A_1) > \util_{v}^{\utilmode}(A_0)$.
    \item  $C_0$ is a winner for~$I^0$, i.e.,
    $C_0$ is feasible for~$I^0$,
    and if $C'$ is feasible for~$I^0$, then $\ScR(C_0)\ge \ScR(C')$.
    \item $A_0$ is a Pareto-$\PBrule$-winning bundle for $I$, i.e.,
      if $A_0\neq C_0$, then $A_0$ $\util^{\utilmode}$-Pareto dominates~$C_0$, and if $C'$ is feasible for $I$ and  $\util^{\utilmode}$-Pareto dominates~$C_0$, then $\ScR(A_0)\ge \ScR(C_0)$.
    \item $A_1$ is a Pareto-$\PBrule$-winning bundle for~$I'$ with $I'=I-\contvec_v+\contvec'_v$, i.e.,
    $A_1 \neq C_0$ and $\util^{\utilmode}$-Pareto dominates~$C_0$,
    and if $C'$~is feasible for $I'$ and $\util^{\utilmode}$-Pareto dominates~$C_0$, then $\ScR(A_1)\ge \ScR(C_0)$.
  \end{compactenum}
     It is fairly straight-forward to verify that the above conditions can be encoded via an \twoqsat{} instance.
  }

  For the sequential variant, the description is a bit more involved. The crucial observation is that since the number of iterations is bounded by~$m+1$.
  To this end, let \myemph{$\cost^*_j\coloneqq \max(0,\cost_j-\sum_{i\in [n]}\contvec_i[j])$}.
  Moreover, for each subset~$X\subseteq C$ of projects, let $\Pot^0(X)$ denote the \pb{} profile derived from~$\Pot$ by restricting to the project subset~$[m]\setminus A_{i-1}$ and without donations.
  
  To decide \probSpendR{\text{\normalfont Sequential-}$\RR$}, we ask whether there exists a contribution vector~$\contvec'_v\in \mathds{N}^m$, $2m+4$~bundles:~$A_0,A_1,\ldots,A_m,C_0,C_1,C_2,\ldots,C_m,A^*,C^*\subseteq [m]$, and a new profile~$Q$ such that for each~$i\in [m]$ for each bundle~$C'\subseteq [m]$ the following hold:
  \begin{compactenum}[(1)]
    \item $\sum_{j\in [m]}\contvec'_v[j] \le \don$.
    \item $Q=\Pot-\contvec_v+\contvec'_v$.
    \item $\util_{v}^{\utilmode}(A^*) > \util_{v}^{\utilmode}(C^*)$.
    \item $A_0=C_0$ and $A_0$ is a winner for~$I^0$, i.e.,
    $A_0$ is feasible for~$I^0$,
    and if $C'$ is feasible for~$I^0$, then $\ScR(A_0)\ge \ScR(C')$.
    \item\label{cond:beforedonation-start} $A_{i-1}\subseteq A_{i}$ and  $A_m\subseteq A^*$. 
    \item If $A_{i-1}\neq A_i$, then $A_i\setminus A_{i-1}$ is a winner for~$I^0_i\coloneqq (\Pot^0([m]\setminus A_{i-1}), B-\sum_{j\in A_{i-1}}\cost^*_j, \lowervec-\sum_{j\in A_{i-1}}\typevec_j, \uppervec-\sum_{j\in A_{i-1}}\typevec_j)$,  i.e.,
    $A_i\setminus A_{i-1}$ is feasible for $I^0_i$ and
    if $C'$ is feasible for~$I^0_i$, then $\ScR(A_i\setminus A_{i-1}) \ge \ScR(C')$.
    \item \label{cond:beforedonation-end} If $A_{i-1} = A_i$,
    then if $C'\subseteq [m]\setminus A_{i-1}$, then $C'$ is not feasible for the instance~$I^0_i$ (see the previous condition).
    \item\label{cond:beforedonation-winner1} If $A_m\neq A^*$, then $A^*\setminus A_m$ is a winner for~$I^*\coloneqq (\Pot([m]\setminus A_{m}), B-\sum_{j\in A_{m}}\cost^*_j, \lowervec-\sum_{j\in A_{m}}\typevec_j, \uppervec-\sum_{j\in A_{m}}\typevec_j)$,  i.e., $A^*\setminus A_m$ is feasible for~$I^*$,
    and if $C'$ is feasible for~$I^*$, then $\ScR(A^*\setminus A_m)\ge \ScR(C')$.
    \item\label{cond:beforedonation-winner2} If $A_m=A^*$, then if $C'\subseteq [m]\setminus A_m$, then $C'$ is not feasible for~$I^*$ (see the previous condition).
    \item\label{cond:afterdonation-start} $C_{i-1}\subseteq C_{i}$ and  $C_{m}\subseteq C^*$.
    \item If $C_{i-1}\neq C_i$, then $C_i\setminus C_{i-1}$ is a winner for~$H^0_i\coloneqq (Q^0([m]\setminus C_{i-1}), B-\sum_{j\in C_{i-1}}\cost^*_i, \lowervec-\sum_{j\in C_{i-1}}\typevec_j, \uppervec-\sum_{j\in C_{i-1}}\typevec_j)$, i.e.,
    $C_i\setminus C_{i-1}$ is feasible for $H^0_i$ and
    if $C'$ is feasible for~$H^0_i$, then $\ScR(C_i\setminus C_{i-1}) \ge \ScR(C')$.
    \item\label{cond:afterdonation-end}  If $C_{i-1} = C_i$,
    then if $C'\subseteq [m]\setminus C_{i-1}$, then $C'$ is not feasible for the instance~$H^0_i$ (see the previous condition).

    \item\label{cond:afterdonation-winner1} If $C_m\neq C^*$, then $C^*\setminus C_m$ is a winner for~$H^*\coloneqq (\Pot([m]\setminus C_{m}), B-\sum_{j\in C_{m}}\cost^*_j, \lowervec-\sum_{j\in C_{m}}\typevec_j, \uppervec-\sum_{j\in C_{m}}\typevec_j)$,  i.e., $C^*\setminus C_m$ is feasible for~$I^*$,
    and if $C'$ is feasible for~$I^*$, then $\ScR(C^*\setminus C_m)\ge \ScR(C')$.
    \item\label{cond:afterdonation-winner2} If $C_m=C^*$, then if $C'\subseteq [m]\setminus C_m$, then $C'$ is not feasible for~$H^*$ (see the previous condition).
  \end{compactenum}
  Note that Conditions~\eqref{cond:beforedonation-start}--\eqref{cond:beforedonation-end} ensure that $(A_0,A_1,\ldots,A_m)$ is the sequence of bundles built in the iterations of the sequential rule for~$I$, while Conditions~\eqref{cond:beforedonation-winner1}--\eqref{cond:beforedonation-winner2} ensure that $A^*$ is a winner for~$I$.
  Conditions~\eqref{cond:afterdonation-start}--\eqref{cond:afterdonation-end}  ensure that $(C_0,C_1,\ldots,C_m)$ is the sequence of bundles built in the iterations of the sequential rule for~$I-\contvec_v+\contvec'_v$, while Conditions~\eqref{cond:afterdonation-winner1}--\eqref{cond:afterdonation-winner2} ensure that $C^*$ is a winner for~$I-\contvec_v+\contvec'_v$.
\end{proof}
\fi

\subsubsection{With no diversity constraints}%

\begin{theorem}\label{thm:donation-notypes:max-sum}
  \probSpendR{$\PBrule^{\max}_{\Sigma}$} is \parallelnp-hard even if there are no diversity constraints and the preferences are dichotomous.
\end{theorem}

\ifshort
\begin{proof}[Proof sketch.]
\else
\begin{proof}
\fi
To show the hardness result, we reduce from a \parallelnp{-complete} problem~\cite[Theorem 3.2.6]{Spakowski2005}, called \maxtruesat:
\iflong
\decprob{\maxtruesat}
  {Two equal-sized sets~$X$ and $Y$ of Boolean variables; two satisfiable Boolean formula~$\phi_1$ and $\phi_2$ over~$X$ and $Y$ in $3^{\le}$CNF of equal size; a $3^{\le}$CNF is a set of clauses each containing at most $3$~literals.}{
    Is it true that $\maxone(\phi_1)\ge \maxone(\phi_2)$?
  }
  \fi
  Herein, given a Boolean formula~$\phi$ over a set~$Z$ of variables, $\maxone(\phi)$ denotes the maximum number of variables set to true in a satisfying truth assignment for $\phi$; if $\phi$ is not satisfiable, then $\maxone(\phi)$ is undefined.
  \iflong
  Since the input of the problem consists of two satisfiable $3^{\le}$CNF, the corresponding $\maxone$ values are well-defined.
  Without loss of generality, we assume that $\phi_1$ and $\phi_2$ each has a satisfying truth assignment where at least one variable is set to true: We can add a new variable and a new clause containing only the new variable to $\phi_1$ and to $\phi_2$ without disturbing the verity of the statement~$\maxone(\phi_1) \ge \maxone(\phi_2)$.
  In the following, we assume \ifshort w.l.o.g.\ \fi that $\maxone(\phi_1) \ge 1$.
  \fi 
  
  The idea of the reduction is to construct, from an instance~$(\phi_1(X),\phi_2(Y))$ of \maxtruesat{} with $|X|=|Y|=\enn$ and $|\phi_1|=|\phi_2|=\emm$,
  an equivalent instance of \probWinnerR{$\PBrule^{\max}_{\Sigma}$}
  with $2\enn$ \myemph{$X$-projects},
  $2\enn$ \myemph{$Y$-projects} (each corresponding to a literal),
  $\enn$ \myemph{auxiliary-projects},
  and $3$ special projects~$x_0, y_0,\alpha_0$, where our target voter~$v$ can only gain satisfaction from~$x_0$.
  We define the costs of the $X$- and $Y$-projects such that projects corresponding to positive literals cost less than projects corresponding to negative literals. Therefore, if more positive projects are funded, then more money is left to select additional auxiliary-projects.
  The costs of the auxiliary-projects are small in comparison to the other projects.
  This way, the score of bundle~$A$ is linear in the number of positive $X$-projects (resp.\ $Y$-projects) in the bundle.
  Moreover, we define the budget, the donation bound, and the costs of~$x_0$ and $y_0$ such that any feasible bundle~$A$ with sufficiently large score satisfies the following properties:
  \begin{inparaenum}[(i)]
    \item $A$ contains either $x_0$ or $y_0$.
    \item If $A$ contains~$x_0$ (resp.\ $y_0$), then it corresponds to a valid truth assignment of $\phi_1$ (resp.\ $\phi_2$).
    \end{inparaenum}      
  Besides voter~$v$, we introduce a large number of additional voters to ensure that any winning (and feasible) bundle must achieve a sufficiently large score. 
  Since voter~$v$ is only satisfied with~$x_0$, the only way for her to improve her utility is to ensure that there exists a winning (and feasible) bundle which includes~$x_0$.
  In order to achieve this, she must donate money to projects such that the number of ``positive'' $X$-projects is at least as large as the number of ``positive'' $Y$-projects in any feasible bundle including~$y_0$, i.e., $\maxone(\phi_1)\ge \maxone(\phi_2)$.

  Formally, let $(\phi_1(X),\phi_2(Y))$ be an instance of \maxtruesat{} with $X=\{x_1,\ldots, x_{\enn}\}$ and $Y=\{y_{1},\ldots,y_{\enn}\}$, $\phi_1=\{C_1,\ldots,C_{\emm}\}$ and $\phi_2=\{D_1,\ldots,D_{\emm}\}$.
  To ease notation, define~\myemph{$L\coloneqq \enn+3$} and \myemph{$R\coloneqq 2\enn+2\emm+4\enn^2+4\enn$}.
  We create an instance of \probSpendR{$\PBrule^{\max}_{\Sigma}$} without types as follows.

  \iflong  \mypa{The projects and their costs.}
  There are three distinguished projects, called~\myemph{$x_0,y_0,\alpha_0$}, with $\cost_{x_0}=2\enn$ and $\cost_{y_0}=\enn$, $\cost_{\alpha_0}=B$; we will set the budget limit~$B$ later.
    For each $X$-variable~$x_i\in X$ we create two \myemph{$X$-projects}, called~$x_i$ and $\overline{x}_i$, associated with~$x_i$.
    Similarly, for each $Y$-variable~$y_i\in Y$ we create two \myemph{$Y$-projects}, called~$y_i$ and $\overline{y}_i$, associated with $y_i$.
    Finally, for each $i\in [\enn]$, we introduce an \myemph{auxiliary-project}, called $\alpha_i$.
    \fi
    The costs of the projects are specified as follows:
    
    {    \centering
    \begin{tabular}{cccccccc}%
      $x_i$ & $\overline{x}_i$ & $y_{i}$ & $\overline{y}_i$ & $\alpha_i$ & $x_0$ & $y_0$ & $\alpha_0$\\ \midrule
      $\enn+1$ & $\enn+2$ & $\enn+1$ & $\enn+2$ & $i$ & $2\enn$ & $\enn$ & $B$%
    \end{tabular}
    \par}

    \smallskip

    The voters have dichotomous preferences, i.e., if they are satisfied with a project, then they are satisfied with value one.
    \ifshort
    Our target voter~$v$ is only satisfied with~$x_0$. %
    Additionally, we create the following $L(2\enn+2\emm+4\enn^2+4\enn)+\enn+5=L \cdot R + \enn + 4$ voters.
    \begin{compactitem}[--]
      \item  For each~$x_i\in X$ (resp.\ $y_i\in Y$), we create $L$ voters~\myemph{$x_i^j$} (resp.\ \myemph{$y_i^j$}), $j\in [L]$, which are only satisfied with the projects $x_i$ and $\overline{x}_i$ (resp.\ $y_i$ and $\overline{y}_i$). %
      \item    For each clause~$C_{\ell} \in \phi_1$ (resp.\ $D_{\ell}\in \phi_2$), we create~$L$ voters~\myemph{$c_{\ell}^{j}$} (resp.\ \myemph{$d_{\ell}^{j}$}), $j\in [L]$, each of which is only satisfied with the $X$-projects (resp.\ $Y$-projects) which correspond to the literals contained in $C_\ell$ (resp.\ $D_{\ell}$), and project~$\alpha_0$. %
      \item    For each~$x_i\in X$, we create~$2\cdot L$~voters~\myemph{$u^j_i$} and \myemph{$\overline{u}^j_{i}$}, $j\in [L]$.
    Each $u^j_i$ (resp.\ $\overline{u}^j_i$) is satisfied with the corresponding $X$-project~$x_i$~(resp.\ $\overline{x}_i$), and projects~$x_0$ and $\alpha_0$. %
    Similarly, for each~$y_i\in Y$, we introduce~$2\cdot L$~voters,~\myemph{$w^j_i$} and \myemph{$\overline{w}^j_{i}$}, $j\in [L]$.
    Each $w^j_i$ (resp.\ $\overline{w}^j_i$) is
    satisfied with the corresponding $Y$-projects~$y_i$~(resp.\ $\overline{y}_i$), and projects~$y_0$ and~$\alpha_0$. %
    
    Finally, for each $\lit\in X\cup \overline{X}$ and $\lit' \in Y \cup \overline{Y}$, we create $L$ connector-voters who are only satisfied with projects~$\lit$, $\lit'$, and $\alpha_0$. %

    \item For each $i\in [\enn]$, we create a voter~\myemph{$a_i$}, who is only satisfied with the auxiliary-projects from $\{\alpha_{\enn},\alpha_{\enn-1}\ldots,\alpha_{\enn-i+1}\}$.
    We create one more voter~\myemph{$a_0$} who is only satisfied with all $\enn$ auxiliary-projects.

    \item Finally, we create three distinguished voters~\myemph{$v_1$}, \myemph{$v_2$}, and \myemph{$v_3$} such that $v_1$ is only satisfied with $y_0$ and $\alpha_0$, while $v_2$ and $v_3$ are only satisfied with $\alpha_0$.
    \end{compactitem}
    \else \mypa{The voters and their satisfactions.}
    In addition to our target voter~$v$, there are five groups of voters, the \myemph{variable-voters}, the \myemph{clause-voters}, the \myemph{connector-voters}, and the \myemph{auxiliary-voters}, the \myemph{distinguished voters}.
All voters except~$v$ and the auxiliary-voters are satisfied with the auxiliary-project~$\alpha_0$.
\begin{description}
  \item[{Target voter~$v$}:] Our target voter is only satisfied with project~$x_0$, and not satisfied with the remaining projects.
  \item[Variable-voters:] There are $2\enn\cdot L$ such voters: For each variable $x_i\in X$, we introduce  $L$ voters, called~\myemph{$x_i^j$}, $j\in [L]$, who each is only satisfied with the $X$-projects~$x_i$ and $\overline{x}_i$, and project~$\alpha_0$, and not satisfied with any other project.
  
Similarly, for each variable $y_i\in Y$, we introduce $L$ voters, called~\myemph{$y_i^j$}, $j\in [L]$, who each is only satisfied with the $Y$-projects~$y_i$ and $\overline{y}_i$, and project~$\alpha_0$, and not satisfied with any other project.
\item[Clause-voters:] There are $2\emm\cdot L$ such voters: For each clause~$C_{\ell} \in \phi_1$ (resp.\ $D_{\ell}\in \phi_2$), we introduce~$L$ voters, called~\myemph{$c_{\ell}^{j}$} (resp.\ \myemph{$d_{\ell}^{j}$}), $j\in [L]$,
who each is only satisfied with the $X$-projects (resp.\ $Y$-projects) which correspond to the literals contained in $C_\ell$ (resp.\ $D_{\ell}$), and project~$\alpha_0$, and not satisfied with any other project.
For instance, if $C_{\ell}=\{x_1,\overline{x}_3,x_4\}$, then voter~$c_{\ell}^{j}$ is satisfied with the $X$-projects~$x_1$, $\overline{x}_3$, $x_4$, and $\alpha_0$.
\item[Connector-voters:] There are $(4\enn^2+4\enn)\cdot L$ such voters. For each variable~$x_i\in X$, we introduce~$2\cdot L$~voters, called~\myemph{$u^j_i$} and \myemph{$\overline{u}^j_{i}$}, $j\in [L]$.
Each $u^j_i$ (resp.\ $\overline{u}^j_i$) is satisfied with the corresponding $X$-project~$x_i$~(resp.\ $\overline{x}_0$), and projects~$x_0$ and $\alpha_0$, and not satisfied with the remaining projects.

Similarly, for each variable~$y_i\in Y$, we introduce~$2\cdot L$~voters, called~\myemph{$w^j_i$} and \myemph{$\overline{w}^j_{i}$}, $j\in [L]$.
Each $w^j_i$ (resp.\ $\overline{w}^j_i$) is satisfied with the corresponding $Y$-projects~$y_i$~(resp.\ $\overline{y}_i$), and projects~$y_0$ and $\alpha_0$, and not satisfied with the remaining projects.

Finally, for each $\lit\in X\cup \overline{X}$ and $\lit' \in Y \cup \overline{Y}$, we introduce $L$ connector-voter, each of whom is only satisfied with projects~$\lit$, $\lit'$, and $\alpha_0$, and not satisfied with the remaining projects.
We use \myemph{$e^{j}_\ell$}, $\ell\in [4\enn^2]$ and $j\in [L]$, to name these connector-voters.
\item[Auxiliary-voters:] There are $\enn+1$ such voters. For each $i\in [\enn]$, we introduce a voter, called~\myemph{$a_i$}, who is only satisfied with the auxiliary-projects from $\{\alpha_{\enn},\alpha_{\enn-1}\ldots,\alpha_{\enn-i+1}\}$.
We additionally introduce a voter~\myemph{$a_0$} who is satisfied with all auxiliary-projects.
\item[Distinguished voters:]
Finally, there are three distinguished voters, called~\myemph{$v_1$}, \myemph{$v_2$}, and \myemph{$v_3$} such that $v_1$ is satisfied with both $y_0$ and $\alpha_0$, while $v_2$ and $v_3$ are only satisfied with~$\alpha_0$. 
\end{description}
In total, we introduced $L(2\enn+2\emm+4\enn^2+4\enn)+\enn+5=L \cdot R + \enn + 5$ voters.
Observe that the preferences of the voters are dichotomous.
\fi

\iflong \mypa{Donations, the budget, and the donation bound.} 
Initially, no voter donates any money to the projects.
The budget~$B$ is set to~$\enn(3\enn+6)$.
This implies that $\{\alpha_0\}$ is a feasible bundle.
The donation bound~$\don$ of voter~$v$ is set to~$\enn$.

This completes the construction of the instance, which can clearly be conducted in polynomial time.
\fi

\ifshort To complete the construction of the instance, we set $B\coloneqq \enn(3\enn+6)$ and $\don\coloneqq \enn$, and let no voter donate any money initially.
Clearly, the construction can be done in polynomial time.
\fi
Note that there are no diversity constraints. %
Let $I$ denote the constructed \pb{} instance.
\ifshort

\newH{\noindent The proof of the following is deferred to an appendix: 

   \noindent (1)~the initial winning bundle has score at least~$L\cdot R+3$ and the initial utility of voter~$v$ is zero, and

   \noindent (2)~$(\phi_1(X),\phi_2(Y))$ and $(I,\don)$ are equivalent, i.e.,
   $\phi_1$ admits a satisfying assignment~$\sigma_1$, such that the number~$k_1$ of $X$-variables set to true is greater or equal to~$\maxone(\phi_2)$ if and only if
   there exists a donation vector for~$v$ with sum at most~$\don$
such that
the feasible bundle consisting of project~$x_0$, the projects corresponding to~$\sigma_1$, all $Y$-projects and project~$\alpha_{k_1}$ is a winner after the donation.}
\else

Before we show the correctness, we observe several properties of feasible bundles.
To this end, %
let $A_0$ denote a winning bundle under~$\PBrule^{\max}_{\Sigma}$.

\begin{claim}\label{cl:donate-no-types-max-sum}
  \begin{compactenum}[(i)]
    \item\label{prop:winning-no-types-max-sum} $\aggsum^{\max}(A_0) \ge L \cdot R+3 =\aggsum^{\max}(\{\alpha_0\})$.
    \item\label{prop:winning-no-types-2-max-sum} For each bundle~$A$ with $\aggsum^{\max}(A) \ge L \cdot R + 3$ it holds that
    each of the variable-, clause-, and the connector-voters must be satisfied with~$A$.
    \item\label{prop:zero-no-types-max-sum} For each bundle~$A'$ with $\cost(A')\le B$ and  $\aggsum^{\max}(A) \ge L\cdot R + 3$ it holds that $\util^{\max}_{v}(A')=0$.
    \end{compactenum}
  \end{claim}

\begin{proof}[Proof of \cref{cl:donate-no-types-max-sum}] \renewcommand{\qedsymbol}{\claimqed}
  To prove Statement~\eqref{prop:winning-no-types-max-sum}, let us consider the bundle~$A'$ containing only~$\alpha_0$.
  It is straight-forward to verify that all voters except the auxiliary-voters and voter~$v$ is satisfied with~$\alpha_0$.
  Since the preferences are dichotomous, it follows that~$\aggsum^{\max}(A') = L\cdot R+3$.
  Hence, any winning bundle under~$\PBrule^{\max}_{\Sigma}$ must have score at least $L\cdot R+3$.

  To show Statement~\eqref{prop:winning-no-types-2-max-sum}, consider an arbitrary bundle~$A$ with $\cost(A)\le B$ and $\aggsum^{\max}(A)\ge L\cdot R+3$.
  Now, observe that for each voter from the variable-, clause-, and  connector-voters there are $L-1$ ``copies'' who have exactly the same preferences as this voter.
  This means that if some voter from the mentioned voter groups would not be satisfied with~$A$, then at least $L-1$ voters are also not satisfied with $A$.
  Consequently, the score of bundle~$A$ is at most $L \cdot (R-1) + \enn + 5 < L \cdot R + 3$, which cannot happen by our assumption; recall that $L=\enn+3$.
  Hence, every voter from the mentioned voter groups is satisfied with~$A$.

  To show Statement~\eqref{prop:zero-no-types-max-sum}, it suffices to show that for each bundle~$A$ with $x_0\in A$ it holds that $\aggsum^{\max}(A)<  L\cdot R+3$.
  Suppose, for the sake of contradiction, that there exists a bundle~$A$ with $x_0\in A$ such that $\aggsum^{\max}(A)\ge L\cdot R+3$.
  By Statement~\eqref{prop:winning-no-types-2-max-sum}, every one of the variable-, clause-, and connector-voters must be satisfied with~$A$.
  Since $x_0\in A$, it follows that $\cost(A\setminus \{x_0\})\le B-2\enn=\enn(3\enn+4)$.
  Hence, $\alpha_0\notin A$.
  Now, we partition $A$ into disjoint subsets as follows:
  $X^*\coloneqq A\cap (X\cup \overline{X})$,
  $Y^*\coloneqq A\cap (Y\cup \overline{Y})$,
  $A^*\coloneqq A\setminus (X^*\cup Y^*\cup \{x_0\})$.  
  
  By the preferences of the $4\enn^2\cdot L$ connector-voters, it must hold that $X\cup \overline{X}\subseteq A$ or $Y\cup \overline{Y}\subseteq A$, i.e.,
  $X^* = X \cup \overline{X}$ or $Y^* = Y\cup \overline{Y}$.
  
  By the preferences of the variable-voters, it follows that for each $x_i\in X$, at least one project from~$\{x_i,\overline{x}_i\}$ must belong to $A$ since otherwise all voters~$x_i^j$, $j\in [L]$ are not satisfied with~$A$ (note that $\alpha_0\notin A$).
  This implies that $|X^*|=\enn$.
  Similarly, for each $y_i\in Y$, at least one project from~$\{y_i,\overline{y}_i\}$ must belong to $A$.
  This implies that $|Y^*|=\enn$.
  
  Altogether, it follows that $\cost(A\setminus \{x_0\})\ge (\enn+1+\enn+2)\cdot \enn + (\enn+1)\cdot \enn = \enn(3\enn+4)$.
  Since $A$ costs at most~$B$, it does no contain any other project from $\{y_0\}\cup \{\alpha_i \mid i \in [\enn]\}$, i.e., $A^* = \emptyset$.
  This implies that none of the three auxiliary voters is satisfied with~$A$.
  Hence, we deduce that $\aggsum^{\max}(A)\le (2\enn+2\emm+4\enn+4\enn^2)\cdot L + 2 < L \cdot R + 3$, a contradiction.  
\end{proof}

By the above claim, we infer that the utility of~$v$ towards any winning bundle under~$\PBrule^{\max}_{\Sigma}$ is zero.
Since $v$ is only satisfied with~$x_0$, to decide whether there is an effective donation for~$v$, we need to decide whether $v$ can donate at most~$\don$ money such that a new winning bundle contains~$x_0$.

\mypa{The correctness.}
We show that %
$\maxone(\phi_1) \ge \maxone(\phi_2)$ is true if and only if
 there exists a donation vector~$b'_v$ and a bundle~$A_1$ such that
\begin{compactenum}[(a)]
\item $x_0\in A_1$, 
\item $\sum{b'_v}\le \don$,
\item $\sum_{j\in A_1}\max(0, \cost_j-b'_v[j])\le B$, and 
\item\label{prop:winning} for all other bundles~$A_2$ with $x_0\notin A_1$ and $\cost(A_2)\le B + \sum_{j\in A_2}b'_v[j]$,
it holds that $\aggsum^{\max}(A_1)\ge \aggsum^{\max}(A_2)$.
\end{compactenum}

For the ``only if'' direction, assume that $\maxone(\phi_1) \ge \maxone(\phi_2)$ is true.
This means that $\phi_1$ and $\phi_2$ are satisfiable and that there exists a satisfying assignment~$\sigma_1$ of $\phi_1$ such that for all satisfying assignment~$\sigma_2$ of $\phi_2$ the number of variables set to true under $\sigma_1$ is greater or equal to that under~$\sigma_2$.
Let $\sigma_1$ be such a witnessing satisfying assignment for the instance, and let $k_1$ be the number of variables set to true under~$\sigma_1$.
Then, $k_1\ge 1$.
Corresponding to~$\sigma_1$, define the following two project sets~\myemph{$X_1\coloneqq \{x_i \in X\mid \sigma_{1}(x_i)=\true\}$} and \myemph{$\overline{X}_1\coloneqq \{\overline{x}_i \in X\mid \sigma_1(x_i)=\false\}$}.

We let voter~$v$ donate all his money ($\delta$) to project~$x_0$, i.e., $b'_v[x_0]=\don$ and let $b'_v[j]=0$ for remaining project~$j$.
Refer the new instance as~$I'$ and define bundle~$A_1$ with \myemph{$A_1=\{x_0\}\cup X_1 \cup \overline{X}_1 \cup Y \cup \overline{Y}\cup \{\alpha_{k_1}\}$}.
We claim that no bundle~$A_2$ with $x_0\notin A_2$ and $\cost(A_2)\le B$ has more score than~$A_1$.
First of all, we calculate the score of $A_1$:
All variable-, \mbox{clause-,} and connector-voters are satisfied with $A_1$.
Moreover, exactly $k_1+1$ auxiliary-voters are satisfied with $A_1$ (due to~$\alpha_{k_1}$).
Hence, $\aggsum^{\max}(A_1) = L\cdot R + 1 + {k_1}+1\ge  L\cdot R + 3$; recall that $k_1\ge 1$.

Suppose, for the sake of contradiction, that there exists another bundle~$A_2$ with $x_0\notin A_2$ and $\cost(A_2)\le B$ such that $\aggsum^{\max}(A_2)>\aggsum^{\max}(A_1)$.
Clearly, it must hold that $\alpha_0\notin A_2$ as otherwise $A_2=\{\alpha_0\}$ and it does not achieve more score than~$A_1$.
Since $A_2$ achieves more score than $A_1$, by \cref{cl:donate-no-types-max-sum}\eqref{prop:winning-no-types-2-max-sum},
it holds that each of the \mbox{variable-}, \mbox{clause-,} and connector-voters must be satisfied with $A_2$.
By the preferences of the $2\enn$ connector voters, it must hold that $X\cup \overline{X}\subseteq A_2$ since $x_0\notin A_2$.
If $y_0\notin A_2$, then it must also hold that $Y\cup \overline{Y} \subseteq A_2$.
However, this will make $A_2$ exceed the budget.
Hence, $y_0\in A_2$.
By the preferences of the $Y$-variable voters, it must hold that for each~$y_i\in Y$,
at least one of the $Y$-projects~$y_i$ and $\overline{y_i}$ belongs to~$A_2$.
If $y_i,\overline{y}_i\in A_2$ for some~$i\in [\enn]$,
then the cost will be
$\cost(A_2)\ge (2\enn+3)\enn+ (\enn+1)\enn + (\enn+2) + \enn > B$.
Hence, for each $y_i\in Y$, exactly one of $\{y_i,\overline{y}_i\}$ belongs to~$A_2$.
In other words, the $Y$-projects in $A_2$ defines a valid truth assignment.
Define \myemph{$Y_1=A_2 \cap Y$} and \myemph{$\overline{Y}_1 = A_2 \cap \overline{Y}$},
and let $\sigma_2$ be a truth assignment such that $\sigma_{2}(y_i)=\true$ if $y_i\in A_2$; and $\sigma(y_i)=\false$ otherwise.
By the $\emm\cdot L$ clause-voters corresponding to $\phi_2$, the defined truth assignment~$\sigma_2$ satisfies~$\phi_2$.
Since $\aggsum^{\max}(A_2)> \aggsum^{\max}(A_1) = L\cdot R + {k_1}+1$,
it follows that $A_2$ contains an auxiliary-project~$\alpha_{k_2}$ such that $k_2 > k_1$.
Moreover, by the cost constraint, $\cost(A_2) = \enn(2\enn+3)+(\enn+2)\enn - |Y_1| + k_2 \le B$.
This implies that $|Y_1|\ge k_2$.
Since $Y_1$ corresponds to exactly the variables which are set to true in $\sigma_2$, it follows that
$\sigma_2$ has a satisfying assignment under which the number of variables set to true is greater than~$k_1$, a contradiction to $(\phi_1,\phi_2)$ being a yes-instance.

For the ``if'' direction, assume that there exists a donation vector~$b'_v$ for~$v$ and a bundle~$A_1$ with $x_0\in A_1$ and $\sum_{j\in A_1}\max(0,\cost_j-\contvec'_v[j]) \le B$ such that
the condition in~\eqref{prop:winning} is satisfied for~$A_1$.
Since $\cost(\alpha_0)+\cost(x_0)=B+2\enn>B+\don$, it follows that $\alpha_0\notin A_1$.
Since $A_0=\{\alpha_0\}$ is also feasible for the new instance and has score at least~$L\cdot R+3$ it follows that $\score(A_1)\ge L\cdot R+3$.
By \cref{cl:donate-no-types-max-sum}\eqref{prop:winning-no-types-2-max-sum},
it follows that each of the variable-, clause-, and connector-voters must be satisfied with~$A_1$.
Since $\alpha_0\notin A_1$, by a reasoning similar to the one for \cref{cl:donate-no-types-max-sum}\eqref{prop:winning-no-types-2-max-sum},
the following hold for~$A_1$:
\begin{compactenum}[(1)]
  \item $X\cup \overline{X}\subseteq A_1$  or  $X\cup \overline{X}\subseteq A_1$;
  \item For each $x_i\in X$, we have $x_i\in A_1$ or $\overline{x}_i \in A_1$;
  \item For each $y_i\in Y$, we have $y_i\in A_1$ or $\overline{y}_i \in A_1$;
  \item\label{cond:exactly-one} If $Y\cup \overline{Y}\subseteq A_1$, then for each $x_i\in X$, either $x_i\in A_1$ or $\overline{x}_i \in A$.
\end{compactenum}
We claim that $y_0\notin A_1$.
Suppose, for the sake of contradiction, that $y_0\in A_1$.
By the costs of the variable-projects, and projects~$x_0$ and~$y_0$,
it follows that $\cost(A_1) \ge (2\enn+3)\enn+(\enn+1)\enn+2\enn+\enn=(3\enn+7)\enn = B+\don$.
This means that $A_1$ contains none of the auxiliary-project.
However, $\aggsum^{\max}(A_1)\le L\cdot R + 2 < \aggsum^{\max}(\{\alpha_0\})$, contradiction to $A_1$ being a winning bundle (for the new \pb{} instance) since $\{\alpha_0\}$ remains feasible for the new \pb{} instance.

Now that $y_0\notin A_1$, by the preferences of $2\enn$ of the connector-voters, it follows that $Y\cup \overline{Y}\subseteq A_1$.
Define $X_1\coloneqq A_1\cap X$ and $\overline{X}_1\coloneqq A_1\cap \overline{X}$.
By the condition in~\eqref{cond:exactly-one}, it follows that $X_1\cup \overline{X}_1$ induce a valid truth assignment for~$\phi_1$.
Let $\sigma_1$ be the induced assignment, i.e., for each variable~$x_i \in X$,
let $\sigma_1(x_i)=\true$ if $x_i\in A_1$; let $\sigma_1(x_i)=\false$ otherwise.
By the preferences of the clause-voters and by our discussion that every clause-voter must be satisfied with~$A_1$ it follows that $\sigma_1$ is a satisfying assignment with $k_1$ variables being set to true, where $k_1=|X_1|$.

We are ready to show that $\maxone(\phi_1) \ge \maxone(\phi_2)$.
Suppose, for the sake of contradiction, that $\maxone(\phi_1)<\maxone(\phi_2)$.
That is, there exists a satisfying truth assignment~$\sigma_2$ of~$\phi_2$ with $k_2$ variables being set to true such that for each satisfying assignment~$\sigma'$ of~$\phi_1$ it holds that the number of variables set to true under~$\sigma'$ is strictly smaller than~$k_2$.
Since~$\sigma_1$ is a satisfying assignment of~$\phi_1$, it follows that $k_1 < k_2$.
Now, define~\myemph{$A_2\coloneqq X\cup \overline{X}\cup Y_1\cup \overline{Y}_1 \cup \{y_0\}\cup \{\alpha_{k_2}\}$}, where $Y_1\coloneqq \{y_i\in Y\mid \sigma_2(y_i)=\true\}$, and $\overline{Y}_1 \coloneqq \{\overline{y}_i\in Y\mid \sigma_2(y_i)=\false\}$. 
One can verify that $A_2$ is feasible since $\cost(A_2)=(2\enn+3)\enn+(\enn+2)\enn-k_2+\enn+k_2=B$.
We claim to obtain a contradiction that $A_2$ achieves more score than~$A_1$.
First of all, it is straight-forward that each of the variable-, clause-, and connector-voters is satisfied with~$A_2$; note that $\sigma_2$ is a satisfying assignment of $\phi_2$.
Moreover, voter~$v_1$ and $k_2+1$ auxiliary-voters are satisfied with $A_1$.
Hence, $\aggsum^{\max}(A_2)=L\cdot R + k_2+2$.
Before we consider the score of~$A_1$ let us recall that $\{x_0\}\cup X_1 \cup \overline{X}_1 \cup Y \cup \overline{Y}\subseteq A_1$ and $\{y_0,\alpha_0\}\cap A_1 =\emptyset$.
Since $\cost(\{x_0\}\cup X_1 \cup \overline{X}_1 \cup Y \cup \overline{Y}) = 2\enn+\enn(\enn+2)-k_1+\enn(2\enn+3)=B+\don-k_1$,
it follows that $A_1$ may contain one of more auxiliary-projects whose overall cost is at most $k_1$. 
By the preferences of the auxiliary-voters, it follows that $\aggsum^{\max}(A_1)\le L\cdot R+2+k_1< L\cdot R+2+k_2=\Sc^{\max}_{\Sigma}(A_2)$, a contradiction.
\fi
\end{proof}

Using an idea similar to the one for~\cref{thm:donation-notypes:max-sum},
we can show the same hardness result for rules using $\agg_{\min}^{\utilmode}$.

\begin{theorem}\label{thm:donation-notypes:*-min}
  For each~$\utilmode\in \{\max, +\}$, \probSpendR{$\PBrule^{\utilmode}_{\min}$} is \parallelnp-hard even if there are no diversity constraints.
\end{theorem}

\iflong\begin{proof}[Proof sketch.]
  The idea is quite similar to the one for \cref{thm:donation-notypes:max-sum}.
  However, to mimic the effect of summation of the utilities in the aggregation,
  we need to use voters which have large cardinal preferences values.
  Given an instance~$(\phi_1(X),\phi_2(Y))$ of \maxtruesat{} with $X=\{x_1,\ldots, x_{\enn}\}$ and $Y=\{y_{1},\ldots,y_{\enn}\}$, $\phi_1=\{C_1,\ldots,C_{{\emm}}\}$ and $\phi_2=\{D_1,\ldots,D_{\emm}\}$.
  Let $C=X\cup \overline{X}\cup Y\cup\overline{Y}\cup\{x_0,y_0,\alpha_0\}\cup \{\alpha_j\mid j \in [\emm]\}$ be the project set which was introduced in the proof of \cref{thm:donation-notypes:max-sum}.
  We first tackle~$\PBrule^{\max}_{\min}$ and then~$\PBrule^{+}_{\min}$.

  \mypa{$\PBrule^{\max}_{\min}$.} The project set~$C$ and their costs remain unchanged.
  
    Besides the target voter~$v$, there are four other groups of voters: the \myemph{variable-}, \myemph{clause-}, \myemph{connector-}, and \myemph{auxiliary-voters}.
    No distinguished voters are needed.
    Moreover, for each group, there will not be multiple copies of the voters with the same preferences.     
    The cardinal preferences also change. Formally:
    \begin{description}
  \item[Target voter~$v$:] Our target voter is satisfied with the three special projects~\myemph{$x_0$}, \myemph{$y_0$}, and \myemph{$\alpha_0$} with values $\enn+1$, $\enn$, and $\enn$, respectively.
  \item[Variable-voters:] There are $2\enn$ such voters: For each variable~$x_i\in X$, we introduce a voter, called~\myemph{$\vx_i$} who is only satisfied with the $X$-projects~$x_i$ and $\overline{x}_i$, and project~$\alpha_0$, with value $\enn$, and not satisfied with any other project.
  Similarly, for each variable $y_i\in Y$, we introduce a voter, called~\myemph{$\vy_i$},
  who is only satisfied with the $Y$-projects~$y_i$ and $\overline{y}_i$, and project~$\alpha_0$, with value $\enn$, and not satisfied with any other project.
\item[Clause-voters:] There are $2\emm$ such voters: For each clause~$C_{\ell} \in \phi_1$ (resp.\ $D_{\ell}\in \phi_2$), we introduce a voter, called~\myemph{$c_{\ell}$} (resp.\ \myemph{$d_{\ell}$}), 
who  is only satisfied with the $X$-projects (resp.\ $Y$-projects) which correspond to the literals contained in $C_\ell$ (resp.\ $D_{\ell}$), and project~$\alpha_0$, with value one, and not satisfied with any other project.
For instance, if $C_{\ell}=\{x_1,\overline{x}_3,x_4\}$, then voter~$c_{\ell}$ is satisfied with the $X$-projects~$x_1$, $\overline{x}_3$, and $x_4$.
\item[Connector-voters:] There are $(4\enn^2+4\enn)$ such voters. For each variable~$x_i\in X$, we introduce~$2$~voters, called~\myemph{$u_i$} and \myemph{$\overline{u}_{i}$},
who is satisfied with the corresponding $X$-project~$x_i$~(resp.\ $\overline{x}_0$), and projects~$x_0$ and $\alpha_0$, with value~$\enn$, and not satisfied with the remaining projects.
Similarly, for each variable~$y_i\in Y$, we introduce $2$~voters, called~\myemph{$w_i$} and \myemph{$\overline{w}_{i}$}, who is satisfied with the corresponding $Y$-projects~$y_i$~(resp.\ $\overline{y}_i$), and projects~$y_0$ and $\alpha_0$, with value $\enn$, and not satisfied with the remaining projects.

Finally, for each $\lit\in X\cup \overline{X}$ and $\lit' \in Y \cup \overline{Y}$, we introduce a connector-voter who is only satisfied with projects~$\lit$, $\lit'$, and $\alpha_0$, with value~$\enn$.
We use \myemph{$e_\ell$}, $\ell\in [4\enn^2]$ to name these connector-voters.
\item[Auxiliary-voter:] There is only one auxiliary voter, called~\myemph{$a$}, and she is satisfied with each auxiliary-project~$\alpha_j$ with value~$j$, and satisfied with $\alpha_0$ with value one. She is not satisfied with any other project.
\end{description}
In total, we introduced $4\enn+2\emm+4\enn^2+4\enn+2$ voters.
The donation and the budget remain the same: no one donates any money initially, and $B\coloneqq \enn(3\enn+6)$.
The donation bound of voter~$v$ remains the same: $\don\coloneqq \enn$.
This completes the construction.

One can obtain a claim similar to~\cref{cl:donate-no-types-max-sum}, where the used score is one instead of $L\cdot R+3$.

\begin{claim}\label{cl:donate-no-types-min}
  \begin{compactenum}[(i)]
    \item\label{prop:winning-no-types} $\aggmin^{\max}(A_0) \ge 1 =\aggmin^{\max}(\{\alpha_0\})$.
    \item\label{prop:winning-no-types-2} For each bundle~$A$ with $\aggmin^{\max}(A) \ge 1$ it holds that
    each of the variable-, clause-, and the connector-voters must be satisfied with~$A$.
    \item\label{prop:zero-no-types} For each bundle~$A'$ with $\cost(A')\le B$ and  $\aggmin^{\max}(A) \ge 1$ it holds that $\util^{\max}_{v}(A')\le \enn$.
    \end{compactenum}
\end{claim}

By a reasoning analogous to the one for~\cref{thm:donation-notypes:max-sum}, using \cref{cl:donate-no-types-min}, one can check that both instances are equivalent.

\mypa{$\PBrule^{+}_{\min}$.} Now, the reduction is almost the same as the one for $\PBrule^{\max}_{\min}$. Hence, we only specify the differences.
Again, the project set remains unchanged.
The costs of the projects except those of the auxiliary projects are unchanged.
Each auxiliary project has unit cost.
  Formally,
  
  {\centering
  \begin{tabular}{@{}|@{\;}c@{\;}|c|c|c|c|@{\;}c@{\;}|c|c|@{\;}c@{\;}|}\toprule
    & $x_i$ & $\overline{x}_i$ & $y_{i}$ & $\overline{y}_i$ & $\alpha_i$ & $x_0$ & $y_0$ & $\alpha_0$\\ \midrule
    cost    & $\enn+1$ & $\enn+2$ & $\enn+1$ & $\enn+2$ & \textcolor{red}{$1$} & $2\enn$ & $\enn$ & $B$\\\bottomrule
  \end{tabular}

  \par}
\smallskip

  The voter set is the same as the one for~$\PBrule^{\max}_{\min}$.
  Their cardinal preferences are the same as those for~$\PBrule^{\max}_{\min}$ except the auxiliary voter~$a$.
  Voter~$a$ is satisfied with each project from~$\{\alpha_j\mid 0\le j \le \enn\}$ with value one, and is not satisfied with the remaining projects.
  In other words, $a$ is still satisfied with the same set of projects as that for~$\PBrule^{\max}_{\min}$, but the cardinal values are different since we use additive utilities instead of maximum.
  The budget bound, initial donations, and the donation bound remain the same.
  
  One can obtain the same statements as in \cref{cl:donate-no-types-min}, by replacing~$\aggmin^{max}$ with $\aggmin^{+}$.
  Using these statements, one can check that both instances are equivalent.
\end{proof}
\fi

\noindent We are not able to show \parallelnp-hardness for rule~$\PBrule^{+}_{\Sigma}$.
  However, we show that it is unlikely to be contained in NP or coNP.
  \begin{theorem}\label{thm:donation-notypes:+-sum}
    \probSpendR{$\PBrule^{+}_{\Sigma}$} is weakly NP-hard and weakly coNP-hard, even if there are no diversity constraints.
  \end{theorem}

\iflong  \begin{proof}
    To show the hardness, we reduce from the weakly NP-complete \knapsack problem and its co-variant, respectively.
    \decprob{\knapsack}{
      $\enn$~items with $2\enn$ positive integers~$s_1,\ldots,s_\enn$ and $v_1,\ldots,v_{\enn}$, and two values~$S$ and $K$.
    }{
      Is there a subset~$J\subseteq [\enn]$ such that $\sum_{j\in J}s_j\le S$ such that $\sum_{j\in J}v_j \ge K$?}

    Let $I=((s_j)_{j\in [\enn]},(v_{j\in [\enn]}), S, K)$ be an instance of \knapsack.
    For technical reason, define~\myemph{$R\coloneqq \sum_{j\in [\enn]}v_j$.}
    We first show NP-hardness, and create an instance of \probSpendR{$\PBrule^{+}_{\Sigma}$} as follows.

    \mypa{The projects and their costs.} For each~$j \in [\enn]$ we introduce one project, called~$x_j$, with costs~$\cost_{x_j}= s_j$. 
    Additionally, we introduce three auxiliary projects~$\alpha_i$, $i\in [3]$, with cost~$S$, $2S$, and $S$, respectively.
    We will set the initial donation in such a way that the initial winning bundle achieves a score of at least~$K+2R$.

    \mypa{The voters and their preferences.}
    Besides the target voter~$v$, there is one more voter, called~$u_1$.
    The target voter is satisfied each project~$x_j$ with value equal to~$v_j$, and with the two auxiliary projects~$\alpha_2$ and $\alpha_3$ with values~$K+R$ and $R+1$, respectively.

    Voter~$u_1$ is satisfied with projects~$\alpha_1$ and $\alpha_2$,
    with values~$K$ and $R$, respectively, and not satisfied with the remaining projects.

    \mypa{The donations, the budget, and the donation bound.}
    The budget bound is set to $B\coloneqq S$.
    Initially, voter~$v$ only made a donation of~$S$ to project~$\alpha_2$, while voter~$u_1$ does not donate any money.
    The donation bound of voter~$v$ is~$\don\coloneqq B = S$.
    \begin{center}\begin{tabular}{@{}|@{\;}l@{\;}|c|c|c|c|}\toprule
      & $x_j$ & $\alpha_1$ & $\alpha_2$ & $\alpha_3$\\\midrule
      cost & $s_j$ & $S$ & $2S$ & $S$\\\hline
      &&&&\\[-1.5ex]
      $\sat_{v}$ & $v_j$ & $0$ & $K+R$ & $R+1$\\\hline
      &&&&\\[-1.5ex]
      $\sat_{u_1}$ & $0$ & $K$ & $R$ & $0$\\\hline 
      &&&&\\[-1.5ex]
     $\contvec_v$ & $0$ & $0$ & $S$ & $0$  \\\hline
      &&&&\\[-1.5ex]
      $\contvec_{u_1}$ & $0$ & $0$ & $0$ & $0$  \\\bottomrule
    \end{tabular}\end{center}    
    This completes the construction.
    Let $I_1$ denote the constructed \pb instance. %
    Before we show the correctness, we observe the following properties of the initial winning bundle for~$I_1$, denoted as~$A_0$.

    \begin{claim}\label{cl:donate-no-types-knapsack}
      \begin{compactenum}[(a)]
        \item $\Sc^{+}_{\Sigma}(A_0)\ge K+2R$.
        \item $A_0=\{\alpha_2\}$.
        \item\label{initial-util} $\util^{+}_v(A_0) = K+R$.
      \end{compactenum}
    \end{claim}

    \begin{proof}
      [Proof of \cref{cl:donate-no-types-knapsack}] \renewcommand{\qedsymbol}{\claimqed}
      To show the first statement, one only need to check that $\{\alpha_2\}$ is a feasible bundle (note that the donation from voter~$v$ to $\alpha_2$ is $S$) and achieves a score of $K+2R$.

      The second statement is also straight-forward since $\alpha_2$ achieves the maximum score  among all feasible bundles.

      The last statement follows from the second statement.
    \end{proof}

    \mypa{The correctness.} We show that $I$ is a yes-instance of \knapsack if and only if $(I_1,\don)$ is a yes-instance of \probSpendR{$\PBrule^{+}_{\Sigma}$}.

    For the ``only if'' direction, assume that $I$ is a yes-instance of \knapsack.
    Let $J \subseteq [\enn]$ denote a subset of indices with $\sum_{j\in J}v_j \le S$ such that $\sum_{j\in J}v_j\ge K$. %
    Then, we let voter~$v$ donate all her money ($S$) to project~$\alpha_3$.
    We show that the following bundle~$A_1$ with $A_1\coloneqq \{x_j\mid j\in J\}\cup \{\alpha_3\}$ is a desired bundle, i.e., it is a winning bundle for the new instance and it makes~$v$ more satisfied.
    First of all, it is straight-forward to check that $A_1$ is feasible.
    Second, $\util^{+}_{v}(A_1) = R+1+\sum_{j\in J}\sat_v(x_j) \ge K+R+1 > \util^{+}_{v}(A_0)$ (see \cref{cl:donate-no-types-knapsack}\eqref{initial-util}).
    Finally, consider an arbitrary feasible bundle~$A_3$ with $\util^{+}_{v}(A_3) \le \util^{+}_{v}(A_0)$.
    We claim that $\Sc^{+}_{\Sigma}(A_3)\le \Sc^{+}_{\Sigma}(A_1) = K+R+1$, which suffices to show the claim. 
    Clearly, $\alpha_2\notin A_3$ since its cost exceeds the budget bound and it receives no donation.
    If $\alpha_1\in A_3$, then by the cost of $\alpha_1$, it follows that $A_3\subseteq \{\alpha_1, \alpha_3\}$ and that $\Sc^{+}_{\Sigma}(A_3)\le K+R+1$.
    If $\alpha_1\notin A_3$, then it follows that $A_3\subseteq \{a_j\mid j\in [\enn]\}\cup\{\alpha_3\}$.
    This means that the score of $A_3$ comes solely from voter~$v$.
    In other words, $\Sc^{+}_{\Sigma}(A_3) = \util_{v}^{+}(A_3)\le K+R$.
    In both cases, we showed that $\Sc^{+}_{v}(A_3)\le K+R+1$. 

    For the ``if'' direction, assume that $(I_1,\don)$ is a yes-instance of  \probSpendR{$\PBrule^{+}_{\Sigma}$}.
    By \cref{cl:donate-no-types-knapsack}, this means that there exists a donation vector~$\contvec'_v$ with $\sum{\contvec'_v}\le \don$ and there exists a winning bundle~$A_1$ for~$I_1-\contvec_v+\contvec'_v$ such that
    $\util^{+}_v(A_1)\ge K+R+1$.
    Clearly, $\alpha_2\notin A_1$ due to its large cost.
    Define $J\coloneqq \{j\in [\enn]\mid a_j\in A_1\}$.
    We claim that $J$ is a witness for $I$ being a yes-instance.
    Suppose, for the sake of contradiction, that $\sum_{j\in J} v_j > S$ or $\sum_{j\in J}v_j < K$.
    First of all, if $\sum_{j\in J}v_j > S$, then it must hold that $\{\alpha_1,\alpha_2,\alpha_3\}\notin A_1$ since otherwise the cost of bundle~$A_1$ exceeds~$B+\don$.
    However, this implies that $\util_{v}^{+}(A_1)\le \util_{v}^{+}(\{a_j\mid j\in [\enn]\}) = R < K+R+1$, a contradiction.
    Hence, we infer that $\sum_{j\in J} v_j \le S$ and $\sum_{j\in J}v_j< K$.
    Since $\alpha_2 \notin A_1$, it follows that $\util_v^{+}(A_1) \le \sum_{j\in J} v_j + R+1< K+R+1$, again a contradiction.

    Now, we turn to the coNP-hardness, and create an instance~$I_2$ with donation bound~$\don_2=1$ as follows.

    \mypa{The projects and their costs.}
    For each~$j \in [\enn]$ we introduce one project, called~$x_j$, with costs~$\cost_{x_j}= s_j$. 
    Additionally, we introduce two auxiliary projects~$\alpha_1$ and $\alpha_2$ with cost~$2S$ and $S+1$, respectively.

    \mypa{The voters and their preferences.}
    Besides the target voter~$v$, there is one more voter, called~$u_1$.
    The target voter is only satisfied with the auxiliary projects~$\alpha_1$ and $\alpha_2$ with values equal to~$K-2$ and $K-1$, respectively.
    The other voter~$u_1$ is satisfied with each project~$x_j$ with value~$v_j$,
    and with project~$\alpha_1$ with value~$R+2$.
    She is not satisfied the other auxiliary projects.

     \mypa{The donations, the budget, and the donation bound.}
    The budget bound is set to $B\coloneqq S$.
    Initially, voter~$v$ only made a donation of~$1$ to each of the auxiliary projects~$\alpha_1$ and $\alpha_2$, while voter~$u_1$
    made a donation of~$S-1$ to project~$\alpha_1$.
    The donation bound of voter~$v$ is~$\don_2\coloneqq 1$.
    \begin{center}
    \begin{tabular}{@{}|@{\;}l@{\;}|c|c|c|}\toprule
      & $x_j$ & $\alpha_1$ & $\alpha_2$ \\\midrule
      cost & $s_j$ & $2S$ & $S+1$\\\hline
      &&&\\[-1.5ex]
      $\sat_{v}$ & $0$ &$K-2$ & $K-1$\\\hline
      &&&\\[-1.5ex]
      $\sat_{u_1}$ & $v_j$ & $R+2$ & $0$\\\hline
      &&&\\[-1.5ex]
      $\contvec_{v}$ & $0$ & $1$ & $1$ \\\hline
      &&&\\[-1.5ex]
      $\contvec_{u_2}$ & $0$ & $S-1$ & $0$\\  \bottomrule
    \end{tabular}
    \end{center}
    This completes the construction.
    Let $I_2$ denote the constructed \pb instance. %
    Before we show the correctness, we observe the following properties of the initial winning bundle for~$I_2$, denoted as~$A'_0$.
    
    \begin{claim}\label{cl:donate-no-types-coknapsack}
      \begin{compactenum}[(a)]
        \item $\Sc^{+}_{\Sigma}(A'_0)\ge R+K$.
        \item $A'_0=\{\alpha_1\}$.
        \item $\util^{+}_v(A'_0) = K-2$.
      \end{compactenum}
    \end{claim}
    \begin{proof}
      [Proof of \cref{cl:donate-no-types-coknapsack}] \renewcommand{\qedsymbol}{\claimqed}
      To show the first statement, one only need to check that $\{\alpha_1\}$ is a feasible bundle  and achieves a score of $R+K$.

      The second statement is also straight-forward since $\alpha_1$ achieves the maximum score among all feasible bundles; recall that $R=\aggsum^{+}(\{x_j\mid j\in [\enn]\})$.

      The last statement follows from the second statement.
    \end{proof}

    \mypa{The correctness.} We show that $I$ is a no-instance of \knapsack if and only if $(I_2,\don_2)$ is a yes-instance of \probSpend{$\PBrule^{+}_{\Sigma}$}.

    For the ``only if'' direction, assume that $I$ is a no-instance of \knapsack.
    Note that, in order to achieve a higher utility for~$v$ we must ensure that $\alpha_2$ is part of a winning bundle since $\sat_v(\alpha_2) = K-1 > K-2$.
    Hence, we let voter~$v$ donate her money to project~$\alpha_2$.
    This makes $\alpha_1$ never part of a feasible bundle since its cost exceeds the budget plus the donation of voter~$u_1$.
    We claim that $A_1=\{\alpha_2\}$ is a winning bundle for the new instance (after $v$ donates~$1$ to project~$\alpha_2$).
    Suppose, for the sake of contradiction, that there exists a feasible bundle~$A_3$ with
    $\Sc^{+}_{\Sigma}(A_3) > \Sc^{+}_{\Sigma}(A_1)$ and  $\util^{+}_{v}(A_3) \le \util^{+}_{v}(A'_0)$.
    This implies that $A_3$ contains no auxiliary project.
    In other words, $A_3 \subseteq \{x_j\mid j\in [\enn]\}$.
    Define $J\coloneqq \{j\in [\enn] \mid x_j\in A_3\}$.
    Then, $\sum_{j\in J}s_j = \cost(A_3)\le B=S$ and $\sum_{j\in J} v_j=\Sc^{+}_{\Sigma}(A_3)\ge K$, a contradiction to $I$ being a no-instance.

    For the ``if'' direction, we show the contra-positive.
    Assume that $I$ is a yes-instance of \knapsack.
    Let $J \subseteq [\enn]$ denote a subset of indices with $\sum_{j\in J}v_j \le S$ such that $\sum_{j\in J}v_j$ is maximum.
    Define bundle $A_3$ with $A_3\coloneqq \{x_j\mid j\in J\}$.
    Observe that $\Sc^{+}_{\Sigma}(A_3) \ge K$.
    Suppose, for the sake of contradiction, that there exists a donation vector which leads to a winning bundle, say~$A_1$, such that the utility of $v$ is more than~$K-2$.
    Since $\alpha_1$ cannot be part of any feasible bundle,
    by the preferences of $v$ it must hold that $\alpha_2\in A_1$.
    By the donation and budget bounds, it follows that $A_1=\{\alpha_1\}$.
    However, $\Sc^{+}_{\Sigma}(A_1) = K-1 < \Sc^{+}_{\Sigma}(A_3)$, a contradiction.
  \end{proof}

\fi

\subsubsection{With diversity constraints}
Using the power of diversity constraints, we can even prove \sigmatwop-hardness.
\iflong That is, finding an effective donation is hard for the complexity class \sigmatwop, whenever diversity constraints are involved. 
\fi
All reductions are from a SAT variant, which is proved to be \sigmatwop-complete by~\citeauthor{ChenGanianHamm2020ijcai-diversestable}~[\citeyear[Claim 1]{ChenGanianHamm2020techreport-diversestable}] and
originally used to prove that finding a diverse and stable matching is \sigmatwop-hard.
\begin{theorem}\label{thm:spend-max-sum-sigma2p-hard}
  \probSpendR{$\PBrule^{\max}_{\Sigma}$} is \sigmatwop-hard, even if the projects have unit costs, the budget is zero, and the preferences are dichotomous.
\end{theorem}

\iflong
\begin{proof}
  To show the hardness for~$\PBrule^{\max}_{\Sigma}$, we provide a polynomial-time reduction from the \sigmatwop-complete \notoneinthree\ problem~\cite{ChenGanianHamm2020ijcai-diversestable} (also see \cite[Claim 1]{ChenGanianHamm2020techreport-diversestable}).
  
  \decprob{\notoneinthree}
{Two equal-sized sets~$X$ and $Y$ of Boolean variables; a Boolean formula~$\phi(X,Y)$ over~$X\cup Y$ in 3CNF, i.e., a set of clauses each containing $3$ literals. Moreover, each clause contains \emph{at least} two literals from $Y\cup \overline{Y}$.}
{Does there exist a truth assignment of $X$ such that for each truth assignment of $Y$ there \emph{exists} a clause~$C_j$ which is not \myemph{1-in-3-satisfied} (i.e., $C_j$ does not have precisely $1$ true literal)?}

The above problem is shown to be \sigmatwop-complete even if each clause contains at least two $Y$-literals~\cite[Claim 1]{ChenGanianHamm2020techreport-diversestable}.

Given a \notoneinthree instance~$\phi(X,Y)$ with $X=\{x_1,x_2,\ldots, x_{\enn}\}$ and $Y=\{y_{\enn+1},y_{\enn+2},\ldots,y_{2\enn}\}$, we create an instance of \probSpendR{$\PBrule^{\max}_{\Sigma}$} as follows.

\mypa{The projects and their costs.} For each $X$-variable~$x_i\in X$ we create two \myemph{$X$-projects}, called~$x_i$ and $\overline{x}_i$, associated with~$x_i$. Similarly, for each $Y$-variable~$y_i\in Y$ we create two \myemph{$Y$-projects}, called~$y_i$ and $\overline{y}_i$, associated with $y_i$.
Further, for each clause~$C_j$, we create a \myemph{clause-project}, called~$z_j$.
Finally, we create four \myemph{auxiliary projects}, called $\alpha_1,\alpha_2,\alpha_3,\alpha_4$ towards each of which our target voter has zero satisfaction.

We will impose budget and diversity constraints in such a way that the bundle consisting of $\alpha_1,\alpha_2$ and $\alpha_3$ is initially a winning bundle since it is the only feasible bundle.
Moreover, a bundle which includes project~$\alpha_4$ represents the situation that $\phi(X,Y)$ is a no-instance of \notoneinthree, i.e., for each truth assignment of $X$ there exists a truth assignment of $Y$ such that each clause in $\phi(X,Y)$ has \emph{exactly} one true literal.

Define~$Z\coloneqq \{z_j\mid j\in [\emm]\}$.
The set of projects is hence~$C\coloneqq X\cup \overline{X} \cup Y\cup \overline{Y}\cup Z \cup \{\alpha_i\mid i\in [4]\}$.

Each project except the four auxiliary projects has a unit cost while the four auxiliary projects cost nothing.

\mypa{The types and the diversity constraints.}
For each clause~$C_j\in \phi(X,Y)$, we create a \myemph{clause-type}, indexed with~$j$. For each $X$-variable~$x_i \in X$ we create an \myemph{$X$-type}, indexed with~$\emm+i$, and similarly, for each $Y$-variable~$y_i \in Y$ we create a \myemph{$Y$-type}, indexed with~$\emm+i$.
Additionally, we create three \myemph{auxiliary types}, indexed with~$\emm+2\enn+1$, ~$\emm+2\enn+2$, and $\emm+2\enn+3$.
These additional types are used to preclude some bundles from being feasible.

\begin{compactitem}[--]
  \item For each literal~$\lit_i\in X\cup \overline{X} \cup Y \cup \overline{Y}$, i.e., $\lit_i\in \{x_i,\overline{x}_i\}$ with $i \in [\enn]$ or $\lit_i \in \{y_i,\overline{y}_i\}$ with $i \in \{\enn+1,\ldots, 2\enn\}$, the type vector of project~$\lit_i$ is defined as:
  \begin{align*}
    & \forall j  \in  [\emm+2\enn+2]\colon\\
    & \typevec_{\lit_i}[j]  \coloneqq
    \begin{cases}
      1,  & \text{if } j = \emm + i \text{ or } \lit_i\!\in C_j\! \text{ for some clasue } C_j,\\
      0, & \text{otherwise.}
    \end{cases}\\
    & \typevec_{\lit_i}[\emm+2\enn+3]  \coloneqq
    \begin{cases}
      1,  & \text{ if } \lit_i \in X \cup \overline{X} ,\\
      0, & \text{ otherwise.}
    \end{cases}
  \end{align*}
  \item For each clause~$C_{i}\in \phi(X,Y)$, the type vector of the corresponding clause-project~$z_i$ is a binary vector with exactly one $1$ at position~$i$, i.e.,
  \begin{align*}
    \typevec_{z_i}[j] \coloneqq
    \begin{cases}
      1,  & \text{ if } j = i, \\
      0, & \text{ otherwise.}
    \end{cases}           
  \end{align*}
  \item The type vectors of the four auxiliary projects are as follows:
  \begin{compactitem}[$\bullet$]
    \item $\typevec_{\alpha_1} \coloneqq 1^{\emm}1^{\enn}1^{\enn}000$.
    \item $\typevec_{\alpha_2} \coloneqq 1^{\emm}0^{\enn}1^{\enn}101$.
    \item $\typevec_{\alpha_3} \coloneqq 1^{\emm}0^{\enn}0^{\enn}011$.
    \item $\typevec_{\alpha_4} \coloneqq  1^{\emm}1^{\enn}1^{\enn}110$.
  \end{compactitem}
\end{compactitem}
The lower- and upper-bounds on the diversity constraints are defined as
$\lowervec\coloneqq 3^{\emm}1^{\enn}2^{\enn}000$ and
$\uppervec\coloneqq 3^{\emm}1^{\enn}2^{\enn}11\enn$.

Note that due to the lower- and upper-bounds on the $X$-types and types~$\emm+2\enn+1$ and $\emm+2\enn+2$, it is straight-forward to verify that no feasible bundle containing project~$\alpha_4$ can include any other auxiliary project~$\alpha_i$ with $i \in [3]$.

\mypa{The voters and their satisfactions.}
In addition to our target voter~$v$, there are $\emm$~\emph{clause-voters}, called $w_k$, $k\in [\emm]$,
and two auxiliary voters, called~$u_1$ and $u_2$.
Our target voter~$v$ is only satisfied with the $X$-projects, with value~$1$, i.e.,
\begin{align*}
  \sat_{v}(x_i) & \coloneqq \sat_{v}(\overline{x}_i) \coloneqq 1, \text{ and }\\
  \sat_{v}(p) & \coloneqq 0 \text{ for each~} p \in C\setminus (X\cup \overline{X}).
\end{align*}
The first auxiliary voter~$u_1$ is satisfied with $\alpha_1$ and $\alpha_4$, each with value~$1$, and not satisfied with the remaining projects.
The second auxiliary voter~$u_2$ is \emph{only} satisfied with $\alpha_4$, with value~$1$.

Each clause-voter~$w_k$, $k\in [\emm]$, is satisfied with only the clause-project~$z_k$ with value~$1$, and not satisfied with any other project.
In total, we introduced $\emm+3$ voters.

\mypa{Donations and the budget.}
No voter donates any money to the projects.
The budget~$B$ is set to zero.

The donation bound~$\don$ of voter~$v$ is set to~$3\enn+\emm$.
In other words, voter~$v$ is free to donate money to any $X/Y$- or clause-project.

\par

This completes the construction of the instance, which can clearly be done in polynomial time.
Let us use~$I$ to refer to the constructed \pb{} instance.
It is straight-forward to verify that $I$ fulfills the restrictions given in the theorem.

First of all, due to the budget bound and the diversity constraints,
it is straight-forward to verify that the initial winning bundle is uniquely $\PBrule^{\max}_{\Sigma}(I) = \{\alpha_1,\alpha_2,\alpha_3\}$.
To ease notation, define~$A_0=\{\alpha_1,\alpha_2,\alpha_3\}$.
Both voters~$u_1$ and $u_2$ have utility one towards~$A_0$ while the remaining $\emm+1$~voters have utility zero towards~$A_0$.
Hence, $\aggsum^{\max}(A_0)=0$.

\mypa{The correctness.}
It remains to show the correctness.
More specifically, we show that there exists a truth assignment~$\sigma_X$ of~$X$ such that for
each truth assignment~$\sigma_Y$ of $Y$ there exists at least one clause~$C_j$ which has either zero, two, or three literals if and only if
donating money to all $Y$-projects, and the projects which correspond to the truth assignment~$\sigma_X$, and some $Z$-projects (so as to meet the constraints on the clause-types) can make them a winning bundle towards which voter~$v$ is more satisfied.

For the ``only if'' direction, assume that $\phi(X,Y)$ is a yes-instance of \notoneinthree and let $\sigma_X$ be a witness for $\phi(X,Y)$ to be a yes-instance, i.e.,
$\sigma_X$ is a truth assignment of~$X$ such that for each truth assignment~$\sigma_Y$ there exits at least one clause~$C_j$ which \emph{does not} have precisely one true literal under~$\sigma_X$ and $\sigma_Y$.

To show that $I$ is also a yes-instance (of \probSpendR{$\PBrule^{\max}_{\Sigma}$}),
let us consider the following bundle~$A_1\coloneqq X'\cup Y \cup \overline{Y} \cup Z'$ with $X'=\{x_i \mid \sigma_X(x_i)=\true\}\cup \{\overline{x}_i \mid \sigma_X(x_i) = \false\}$ and $Z'=\{z_j \mid |C_j\cap (X'\cup Y\cup \overline{Y})| = 2\}$.
Recall that the original donation vector~$\contvec_v$ of~$v$ is an all-zero vector.
We let the new donation vector~$\contvec'_v$ of~$v$ be one for each project from $A_1$ and zero for each remaining project.
Define the new \pb{} instance $I'\coloneqq I-\contvec_v+\contvec'_v$.

One can verify that $A_1$ is feasible for the new instance~$I'$ due to the following:
\begin{compactitem}[--]
  \item $\cost(A_1) = \sum_{j\in A_1}\contvec'_v[j]$,
  \item $A_1$ satisfies the diversity constraints corresponding to the $X$-, $Y$-, and the auxiliary types.
  \item It remains to consider the diversity constraints of the clause-types. Recall that by the definition of \notoneinthree{}, each clause~$C_j\in \phi(X,Y)$ contains at least two $Y$-literals.
  In other words, for each clause-type~$j\in [\emm]$ it holds that $\sum_{p\in X'\cup Y \cup \overline{Y}}\typevec_p[j] \ge 2$.
  By the definition of~$Z'$, it is immediate that the diversity constraints of each clause-type is satisfied.
\end{compactitem}
As for the utility of voter~$v$ and the score of $A_1$, clearly, $\util^{\max}_{v}(A_1) = 1 > \util_{v}(A_2)$ and the score is $\Sc^{\max}_{\Sigma}(A_1) = 1+|Z'|$; observe that for each clause-type~$z_j\in Z'$ the corresponding clause-voter has utility one towards~$A_1$.
It remains to show that $A_1$ is a winning bundle of the new \pb{} instance~$I'$.
In other words, for each feasible bundle~$A_3\subseteq C$ in the new \pb{} instance~$I'$ it holds that
\begin{compactitem}[--]
  \item $\Sc^{\max}_{\Sigma}(A_3) \le \Sc^{\max}_{\Sigma}(A_1)$, or
  \item $\util_{v}^{\max}(A_3)\ge \util^{\max}_{v}(A_1)$.
\end{compactitem}
Suppose, for the sake of contradiction, that there exists a feasible bundle~$A_3$ with $\Sc^{\max}_{\Sigma}(A_3) > \Sc^{\max}_{\Sigma}(A_1)$ and $\util_{v}^{\max}(A_3) < \util^{\max}_{v}(A_1)$.
By the preferences of~$v$, it follows that $(X\cup \overline{X}) \cap A_3 = \emptyset$.
By the constraints on the $X$-types, it follows that either $\alpha_1\in A_3$ or $\alpha_4\in A_3$.
By the score assumption, it follows that $\Sc^{\max}_{\Sigma}(A_3) \ge |Z'|+2$.
However, since no other $Z$-projects can gain any donation from~$v$ and since $B=0$ it follows that the only possible way to make $A_3$ achieve a score of at least~$|Z'|+2$ is to let $A_3 \supseteq \{a_4\}\cup Z'$.
By the upper-bounds on the auxiliary types~$\emm+2\enn+1$ and~$\emm+2\enn+2$, it follows that $\{\alpha_2, \alpha_3\}\cap A_3 = \emptyset$.
By the constraints on the $Y$-types, it follows that for each $y_i\in Y$, the defeater~$A_3$ includes either $y_i$ or $\overline{y}_i$.
Summarizing, %
\begin{align*}
  A_3 & = \{\alpha_4\} \cup Z' \cup Y_1 \text{ such that }\\
  |Y_1| & = \enn \text{ and } |Y_1\cap \{y_i,\overline{y}_i\}| = 1 \text{ for each } y_i \in Y.
\end{align*}
In other words, both $Y_1$ and $(Y\cup \overline{Y}) \setminus Y_1$ correspond to truth assignments of $Y$ which are complementary to each other.
Define $Y_2 \coloneqq (Y\cup \overline{Y}) \setminus Y_1$ and
let $\sigma_Y$ be a truth assignment corresponding to~$Y_2$, i.e., $\sigma_Y(y_i)=\true$ if $y_i\in Y_2$ and $\sigma_Y(y_i)=Y_2$ otherwise.
Recall that $\phi(X,Y)$ is a yes-instance and $\sigma_X$ is such a witness.
Hence, for~$\sigma_Y$, there exists a clause~$C_j$ which does not have exactly one true literal under~$\sigma_X$ and $\sigma_Y$.
By the constraints of the corresponding clause-type~$j$ and since $\typevec_{\alpha_4}[j]=1$,
it follows that $Z'\cup Y_1$ contains two projects which have type~$j$.
We distinguish between two cases, aiming to deduce that 
$|C_j \cap (X'\cap Y_2)| = 1$.
\begin{compactenum}%
  \item[Case 1: $z_j\in Z'$.] Then, by construction, it follows that $|C_j\cap (X'\cup Y\cup \overline{Y})| = 2$.
  Moreover, %
  $Y_1$ contains exactly one project which has type~$j$, %
  i.e., $|C_j\cap Y_1|=1$.
  This implies that $|C_j \cap (X'\cap Y_2)| = 1$.
  \item[Case 2: $z_j \notin Z'$.] Then, by construction, it follows that $|C_j \cap (X'\cup Y \cup \overline{Y})| = 3$.
  Moreover, $Y_1$ must contain exactly two  projects with type~$j$,
  i.e., $|C_j \cap Y_1|=2$.
  This also implies that $|C_j \cap (X'\cap Y_2)| = 1$.
\end{compactenum}
In both cases, we showed that $|C_j \cap (X'\cap Y_2)| = 1$.
This implies that $C_j$ has exactly one true literal under~$\sigma_X$ and $\sigma_Y$, a contradiction.
Hence,~$A_1$ is indeed a winning bundle for the new \pb{} instance, for which voter~$v$'s utility is larger (than the one for~$A_0$).

For the ``if'' direction, assume that $I$ is a yes-instance of \probSpendR{$\PBrule^{\max}_{\Sigma}$} and let $\contvec'_v$ be a donation vector with $\sum{\contvec'_b}\le \don$ and let $A_1$ be a winning bundle under~$I'=I-\contvec_v+\contvec'_v$ with $\util^{\max}_{v}(A_1) > \util^{\max}_{v}(A_2)=0$, where $\contvec_{v}$ is an all-zero donation vector of~$v$ in instance~$I$.
By the preferences of~$v$, it follows that $A_1$ contains at least one $X$-project.
By the constraints on the $X$-types, it follows that $A_1 \cap \{\alpha_1,\alpha_4\}=\emptyset$.
Define $X'\coloneqq A_1\cap (X\cup \overline{X})$.
Then, by the same constraints on the $X$-types, for each $x_i\in X$ it holds that $|X'\cap \{x_i,\overline{x}_i\}|=1$.
By the upper-bound of type~$\emm+2\enn+3$, it follows that $A_1\cap \{\alpha_2,\alpha_3\}=\emptyset$.
By the constraints of the $Y$-types, it follows that $(Y\cup \overline{Y}) \subseteq A_1$.
By the constraints of the clause-types, it follows $A_1\cap Z=\{z_j \mid |C_j\cap (X'\cup Y\cup \overline{Y})|=2\}$.
Define $Z'=\{z_j \in Z \mid |C_j\cap (X'\cup Y\cup \overline{Y})|=2\}$.
Then, we have $A_1=X'\cup Y\cup \overline{Y}\cup Z'$.
By the cost of the projects in~$A_1$ it follows that $\contvec_v[p]=1$ for each $p\in A_1$.

Now, to show that $\phi(X,Y)$ is a yes-instance of \notoneinthree{}, we first observe that $X'$ corresponds to a truth assignment of $X$.
Define the corresponding truth assignment~$\sigma_X$, i.e., for each~$x_i \in X$, define $\sigma_{X}(x_i)\coloneqq \true$ if $x_i \in X'$; $\sigma(x_i)=\false$ otherwise.
We claim that for each truth assignment~$\sigma_Y$ of $Y$ there exists a clause~$C_j$ which does not have exactly one true literal under~$\sigma_X$ and $\sigma_Y$.
To this end, let $\sigma_Y$ be an arbitrary truth assignment of~$Y$.
Let $Y_2$ be the set of projects corresponding to~$\sigma_Y$, i.e., $Y_2 \coloneqq \{y_i \mid \sigma_Y(y_i)=\true\}\cup \{\overline{y}_i \mid \sigma_Y(y_i)=\false\}$ and let $Y_1$ be its complement, i.e., $Y_1\coloneqq (Y\cup \overline{Y})\setminus Y_2$.
Consider the following bundle~$A_3$ with $A_3\coloneqq Y_1 \cup Z' \cup \{\alpha_4\}$.
Clearly, $\util^{\max}_v(A_3) = 0 < \util^{\max}_v(A_1)$ and $\Sc^{\max}_{\Sigma}(A_3) = |Z'| + 2 > \Sc^{\max}_{\Sigma}(A_1)$.
However, since $A_1$ is supposed to be the winning bundle, it must hold that~$A_3$ violates one of the diversity constraints.
Since $A_3$ satisfies the diversity constraint of the last $2\enn+3$~types, there exists a clause-type~$j\in [\emm]$ such that $\sum_{p\in A_3}\typevec_{p}[j] \neq 3$.
We claim that clause~$C_j$ fulfills the condition that we are searching for, i.e., it does not have exactly one true literal under~$\sigma_X$ and $\sigma_Y$.
Since $\typevec_{\alpha_4}[j] = 1$, it further implies that
\begin{align}\label{eq:specialY1}
  \sum_{p\in Y_1\cup Z'}\typevec_{p}[j] \neq 2.
\end{align}
We distinguish between two cases.
\begin{compactenum}
  \item[Case 1: $z_j \in Z'$.] Then, by \eqref{eq:specialY1}, we have that
  $|Y_1 \cap C_j| \neq 1$.
  Moreover, by our construction of~$Z'$ we also infer that $|C_j \cap (X'\cup Y \cup \overline{Y})|=2$.
  This means that $|C_j \cap (X'\cup Y_2)| \neq 1$.
  \item[Case 2: $z_j \notin Z'$.] Then, by \eqref{eq:specialY1}, we have that
  $|Y_1 \cap C_j| \neq 2$.
  Moreover, by our construction of~$Z'$ we also infer that $|C_j \cap (X'\cup Y \cup \overline{Y})|=3$.
  This means that $|C_j \cap (X'\cup Y_2)| \neq 1$.
\end{compactenum}
In both cases, since $\sigma_X$ and $\sigma_Y$ correspond to the sets~$X'$ and $Y_2$, respectively,
we further infer that $C_j$ is a clause which does not have a true literal under~$\sigma_X$ and $\sigma_Y$, as desired.
Since $\sigma_Y$ is an arbitrary truth assignment, we show that $\phi(X,Y)$ is a yes-instance of \notoneinthree.
\end{proof}
\fi
The reduction in the proof of \cref{thm:spend-max-sum-sigma2p-hard} can be extended to show the same hardness for $\PBrule^{+}_{\Sigma}$.

\begin{corollary}\label{cor:spend-+-sum-sigma2p-hard}
   \probSpendR{$\PBrule^{+}_{\Sigma}$} is \sigmatwop-hard, even if the projects have unit costs, and the budget is zero, and the preferences are dichotomous.
 \end{corollary}

 \iflong
 \begin{proof}[Proof sketch.]
   The reduction is almost the same as the one for \cref{thm:spend-max-sum-sigma2p-hard}.
   The only difference lies in the voter set since we need to adjust for the utility function~$\util^{+}$.
   More precisely, we add $\enn-2$ voters who have the same satisfaction as voter~$u_1$
   and one more voter who has the same satisfaction as voter~$u_2$.
   Neither do these newly added voters donate any money to the projects.

   It is straight-forward to verify that $A_2=\{\alpha_1,\alpha_2,\alpha_3\}$ is still the initial winning bundle, with score~$\enn-1$.
   The winning bundle~$A_1$ which corresponds to a witness~$\sigma_X$ of a yes-instance of \notoneinthree{}
   remains the same and has score~$\enn+|Z'|$, where $Z'$ is defined as in ``only if'' direction in the proof of \cref{thm:spend-max-sum-sigma2p-hard}.
   By the satisfactions of the newly added voters and the two auxiliary voters, any bundle which could defeat $A_1$ must include $\alpha_4$ and exclude all projects from $X\cup \overline{X}$.
   Using the above observations, we can show the correctness of the construction.
 \end{proof}
\fi

Again, we can adapt the reduction for \cref{thm:spend-max-sum-sigma2p-hard} to show hardness for aggregation rules which maximize the minimum utility of all voters.

\begin{proposition}\label{prop:spend-max-min-sigma2p-hard}
  \probSpendR{$\PBrule^{+}_{\min}$} (resp.\ \probSpendR{$\PBrule^{\max}_{\min}$}) is \sigmatwop-hard, even if the projects have unit costs, and the budget is zero, and
  the preferences are dichotomous (resp.\ trichotomous).
 \end{proposition}

\iflong \begin{proof}[Proof sketch.]
   We first tackle~$\PBrule^{+}_{\min}$, and then~$\PBrule^{\max}_{\min}$.
   Both reductions are from \notoneinthree{} and similar to the one for \cref{thm:spend-max-sum-sigma2p-hard}.
   Let $\phi(X,Y)$ be an instance of \notoneinthree{} with $|X|=|Y|=\enn$ and $|\phi(X,Y)|=\emm$,
   and let $I$ denote the \pb{} instance created in the proof of \cref{thm:spend-max-sum-sigma2p-hard}. %
   We modify instance~$I$ to show the hardness for~$\PBrule^{\max}_{\min}$. 
   The differences lie in the project set and the satisfactions of some of the voters. %

   \mypa{The projects and their costs.}
   Recall that in the proof for \cref{thm:spend-max-sum-sigma2p-hard},
   we introduced one clause-project~$z_j$ and one clause-voter~$w_j$ for each clause~$C_j\in \phi(X,Y)$.
   Now, we introduce an additional copy of each clause-project~$z_j$, called $d_j$.
   Define~$D\coloneqq \{d_j \mid j\in [\emm]\}$.
   The only difference is that each of the newly introduced clause-project~$d_j$ does not possess any type but clause-voter~$w_j$ is also satisfied with it.
   Together, the project set is $C=X\cup \overline{X}\cup Y \cup \overline{Y}\cup Z \cup D \cup \{\alpha_i \mid i \in [4]\}$.
   
   The costs of the projects from $C\setminus D$ remain the same,
   and each project from $D$ has unit cost.

   \mypa{The types and the diversity constraints.}
   The types of the projects from $C\setminus D$ remain the same,
   while no project from $D$ possesses any type.
   The diversity constraints remain the same.

   \mypa{The voters and their satisfactions.}
   The voter set remains the same.
   We have the following adjustment of the satisfaction functions. 
   Recall that in the proof for \cref{thm:spend-max-sum-sigma2p-hard}
   we introduced, in addition to our target voter~$v$, $\emm$~\emph{clause-voters}, called $w_j$, $j\in [\emm]$,
   and two auxiliary voters, called~$u_1$ and $u_2$.
   Our target voter~$v$ is now satisfied with each of the $X$-projects and auxiliary projects~$\alpha_1$ and $\alpha_4$, each with value~$1$, i.e.,
   \begin{align*}
     \sat_{v}(x_i) & \coloneqq \sat_{v}(\overline{x}_i) \coloneqq 1,\\
     \sat_v(\alpha_1) & \coloneqq \sat_v(\alpha_4) \coloneqq 1, \text{ and }\\
     \sat_{v}(p) & \coloneqq 0 \text{ for each~} p \in C\setminus (X\cup \overline{X} \cup \{\alpha_1,\alpha_4\}).
   \end{align*}
   Each clause-voter~$w_j$, $j\in [\emm]$, is satisfied with the clause-project~$z_j$ and its copy~$d_j$, each with value~$1$, 
   and \emph{not} satisfied with the remaining projects.
   The satisfactions of the auxiliary voters remain the same:
   The first auxiliary voter~$u_1$ is satisfied with $\alpha_1$ and $\alpha_4$, each with value~$1$, and not satisfied with the remaining projects.
   The second auxiliary voter~$u_2$ is \emph{only} satisfied with $\alpha_4$, with value~$1$.

   \mypa{Donations, the budget, and the donation bound.} The donations, the budget, and the donation bound remain the same. That is, no voter donates any money initially, $B\coloneqq 0$, and $\don \coloneqq 3\enn+\emm$.

   This completes the construction of the instance, which can clearly be done in polynomial time.
   Let us use~$I$ to refer to the constructed \pb{} instance.
   It is straight-forward to verify that $I$ fulfills the restrictions given in the theorem.

   \mypa{The correctness.}
   It is straight-forward to verify that $A_0=\{\alpha_1,\alpha_2,\alpha_3\}$ is an initial winning bundle since it is the only feasible bundle.
   This winning bundle has score~$0$ since for instance voter $u_2$ has zero utility towards it.
   The target voter~$v$, however, has utility one towards~$A_0$, i.e., $\util^+_v(A_0)=1$; note that the additive utility is used.

   Let $I'$ denote the modified instance.
   We show that $\phi(X,Y)$ is a yes-instance if and only if $I'$ is a yes-instance of \probSpendR{$\PBrule^{+}_{\min}$}.

   For the ``only if'' direction, assume that $\phi(X,Y)$ is a yes-instance and let $\sigma_{X}$ be a witness for this.
   We claim that if voter~$v$ donates money to each project from the bundle~$A_1$ with $A_1\coloneqq$$X'\cup Y\cup \overline{Y}\cup Z' \cup D'$ with
   $X'\coloneqq \{x_i \in X \mid \sigma_X(x_i)=\true\} \cup \{\overline{x}_i \in \overline{X} \mid \sigma_X(x_i)=\false\}$,
   $Z'\coloneqq \{z_j \in Z\mid |C_j\cap (X'\cup Y\cup \overline{Y})| = 2\}$, and
   $D'\coloneqq \{d_j \in D \mid z_j \in Z\setminus Z'\}$,
   then $A_1$ is a winning bundle under the new instance with the donations towards which $v$ has a higher utility than towards~$A_0$; without loss of generality we assume that $\enn > 1$.
   Clearly, $\cost(A_1)=\don$.
   It is straight-forward to verify that $v$ has utility $\enn$ towards~$A_1$ which is higher than to~$A_0$,
   while each clause-voter~$w_j$ has utility one towards~$A_1$, due to either~$z_j$ or~$d_j$.
   It is also straight-forward to verify that $A_1$ is feasible and has score~$0$ since for instance, no auxiliary voter is satisfied with~$A_1$.
   It remains to show that $A_1$ is a desired winning bundle.
   Suppose, for the sake of contradiction, that there exists another feasible bundle~$A_3$ with
   \begin{compactitem}[--]
     \item $\Sc^{+}_{\min}(A_3) > \Sc^{+}_{\min}(A_1)$ and
     \item $\util_{v}^{+}(A_3) \le \util^{+}_{v}(A_0)$.
   \end{compactitem}
   By the above utility relation and by the preferences of~$v$, this means that $|X\cup \overline{X}) \cap A_3|\le 1$.
   By the diversity constraint on the $X$-type~$x_i$, it follows that either $\alpha_1\in A_3$ or $\alpha_4\in A_3$.
   By the above score inequality, it follows that $\Sc^{+}_{\min}(A_3) > \Sc^{+}_{\min}(A_1)=0$.
   This means that all clause-voters and the second auxiliary voter~$u_2$ must each have utility at least one towards~$A_3$.
   Since $u_2$ is only satisfied with $\alpha_4$, it follows that $\alpha_4\in A_3$.
   Since no other $Z$-projects or $D$-projects can gain any donation from~$v$ and since $B=0$ it follows that the only possible way to make $A_3$ achieve a score of at least~$1$ is to ensure that $A_3 \supseteq \{a_4\}\cup Z' \cup D'$.
   
   By the upper-bounds on the auxiliary types~$\emm+2\enn+1$ and~$\emm+2\enn+2$, it follows that $\{\alpha_2, \alpha_3\}\cap A_3 = \emptyset$.
   By the constraints on the $Y$-types, it follows that for each $y_i\in Y$, the defeater~$A_3$ includes either $y_i$ or $\overline{y}_i$.
   The remaining proof follows exactly the same way as the one for \cref{thm:spend-max-sum-sigma2p-hard}.
   
   For the ``if'' direction, assume that $I'$ is a yes-instance of \probSpendR{$\PBrule^{+}_{\min}$} and let $\contvec'_v$ be a donation vector with $\sum{\contvec'_v}\le \don$ and let $A_1$ be a winning bundle under~$I''=I'-\contvec_v+\contvec'_v$ with $\util^{+}_{v}(A_1) > \util^{+}_{v}(A_0)=1$, where $\contvec_{v}$ is an all-zero donation vector of~$v$ in instance~$I'$.
   Since $\alpha_1$ and $\alpha_4$ cannot be both included to~$A_1$ due to the diversity constraints on the $X$-types, by the preferences of~$v$,
   it follows that $A_1$ contains at least one $X$-project.
   Again, by the constraints on the $X$-types, it follows that $A_1 \cap \{\alpha_1,\alpha_4\}=\emptyset$.
Define $X'\coloneqq A\cap (X\cup \overline{X})$.
Then, by the same constraints on the $X$-types, for each $x_i\in X$ it holds that $|X'\cap \{x_i,\overline{x}_i\}|=1$.
By the upper-bound of type~$\emm+2\enn+3$, it follows that $A_1\cap \{\alpha_2,\alpha_3\}=\emptyset$.
By the constraints of the $Y$-types, it follows that $(Y\cup \overline{Y}) \subseteq A_1$.
By the constraints of the clause-types,
for each clause-project~$z_j\in Z$ it holds that $z_j \in A_1$ if and only if
$|C_j\cap (X'\cup Y\cup \overline{Y})|=2$.
Define $Z'\coloneqq \{z_j \in Z \mid |C_j\cap (X'\cup Y\cup \overline{Y})|=2\}$
and $D'\coloneqq \{d_j \in D \mid z_j \in Z\setminus Z'\}$.
Then, we have $A_1=X'\cup Y\cup \overline{Y}\cup Z'\cup D'$.
Observe that both auxiliary voters have zero utility towards $A_1$.
By the cost of the projects in~$A_1$ it follows that $\contvec_v[p]=1$ for each  project~$p\in A_1$.

Now, to show that $\phi(X,Y)$ is a yes-instance of \notoneinthree{}, we first observe that $X'$ corresponds to a truth assignment of $X$.
Define the corresponding truth assignment~$\sigma_X$, i.e., for each~$x_i \in X$, define $\sigma_{X}(x_i)\coloneqq \true$ if $x_i \in X'$; $\sigma(x_i)=\false$ otherwise.
We claim that for each truth assignment~$\sigma_Y$ of $Y$ there exists a clause~$C_j$ which does not have exactly one true literal under~$\sigma_X$ and $\sigma_Y$.
To this end, let $\sigma_Y$ be an arbitrary truth assignment of~$Y$.
Let $Y_2$ be the set of projects corresponding to~$\sigma_Y$, i.e., $Y_2 \coloneqq \{y_i \mid \sigma_Y(y_i)=\true\}\cup \{\overline{y}_i \mid \sigma_Y(y_i)=\false\}$ and let $Y_1$ be its complement, i.e., $Y_1\coloneqq (Y\cup \overline{Y})\setminus Y_2$.
Consider the following bundle~$A_3$ with $A_3\coloneqq Y_1 \cup Z' \cup D' \cup \{\alpha_4\}$.
One can verify that $\util^{+}_v(A_3) = 1 \le \util^{+}_v(A_0)$ and $\Sc^{+}_{\min}(A_3) = 1 > 0 =\Sc^{+}_{\min}(A_1)$.
To prevent $A_3$ from being a winner, it must hold that $A_3$ violates one of the diversity constraints.
Since $A_3$ satisfies the diversity constraints on the last $2\enn+3$~types, there exists a clause-type~$j\in [\emm]$ such that $\sum_{p\in A_3}\typevec_{p}[j] \neq 3$.
The remaining proof that clause~$C_j$ fulfills the condition that we are searching for, i.e., it does not have exactly one true literal under~$\sigma_X$ and $\sigma_Y$ follows that same way as the one for \cref{thm:spend-max-sum-sigma2p-hard}.

Finally, we turn to $\PBrule^{\max}_{\min}$.
The reduction is almost the same as the one given above.
The only difference lies in that the target voter~$v$ has cardinal preference of value two instead of one towards each $X$-project.
Formally,
  \begin{align*}
     \sat_{v}(x_i) & \coloneqq \sat_{v}(\overline{x}_i) \coloneqq 2,\\
     \sat_v(\alpha_1) & \coloneqq \sat_v(\alpha_4) \coloneqq 1, \text{ and }\\
     \sat_{v}(p) & \coloneqq 0 \text{ for each~} p \in C\setminus (X\cup \overline{X} \cup \{\alpha_1,\alpha_4\}).
  \end{align*}
  Since $\util^{\max}$ is used, an optimal donation must result in a winning bundle which contains at least one $X$-project.
  The correctness proof follows in the same manner.
 \end{proof}
 
\fi

\iflong
\paragraph{The Sequential- and Pareto-variants.}

In order to show \sigmatwop-hardness for these aggregation variants,
we observe that if we modify the instances constructed in the proofs for \cref{thm:spend-max-sum-sigma2p-hard}, \cref{cor:spend-+-sum-sigma2p-hard}, and \cref{prop:spend-max-min-sigma2p-hard}  so that all four auxiliary projects have unit costs and the auxiliary voter~$v_2$ donates to each of the auxiliary projects that amount of money,
the modified  instances can be used to directly show \sigmatwop-hardness for the sequential variants.
Moreover, for each of the considered feasible bundles~$A_1,A_2$ since its Pareto-dominating set is of size one, it is also fairly straight-forward to see that the same reduction works for the Pareto-valiant.
\fi

\begin{corollary}
  For each $\PBrule \in \aggSet$, %
  and each variant~$\phi\in \{$Sequential, Pareto$\}$,
  \probSpendR{$\phi$-$\PBrule$} is \sigmatwop-hard.
\end{corollary}

\section{Outlook}

To conclude, we would like to mention that greedy rules, which are often used in
practice, easily fit our framework (although they might fail to respect diversity lower bounds). 
In particular the combination of a greedy rule~$\PBrule$ with the Sequential-$\PBrule$ method to handle donations is promising; preliminary research shows that such a rule satisfies donation-no-harm
and can be computed in polynomial-time.
In general, the axiomatic and computational analysis of PB with donations requires further attention before a comprehensive assessment on its usefulness can be given.

\section*{Acknowledgments}
Jiehua Chen is supported by the WWTF research project~(VRG18-012).
Martin Lackner and Jan Maly are supported by the FWF research project~P31890.

\clearpage
\bibliographystyle{named}
\bibliography{bib}

\appendix

\end{document}